\documentclass[12pt]{article}
\usepackage[lmargin=1.in,rmargin=1.in,tmargin=1.in,bmargin=1in]{geometry}
\usepackage[utf8]{inputenc}
\usepackage[english]{babel}
\usepackage{geometry}
\usepackage{gensymb}
\usepackage{textcomp}
\usepackage{hyperref}
\usepackage{floatrow}
\usepackage[label font=bf,labelformat=simple]{subfig}
\usepackage{caption}
\usepackage{adjustbox,lipsum}
\usepackage[autostyle]{csquotes}
\usepackage{dirtytalk}
\usepackage{amsmath}
\usepackage{tikz}
\usepackage[T1]{fontenc}
\usetikzlibrary{matrix,shapes,arrows,positioning,chains}
\usepackage{array,multirow,graphicx}
\usepackage{authblk}
\usepackage{hyperref}
\usepackage[font=footnotesize]{caption}
\usepackage[
backend=bibtex,
style=numeric-comp,
natbib=true,
uniquelist=false,
sorting=none,maxcitenames=1,maxbibnames=8
]{biblatex}
 \usepackage[document]{ragged2e}
 \providecommand{\keywords}[1]
{
  \small	
  \textbf{\textit{Keywords---}} #1
}
\title{Transoceanic migration of dragonflies and branched optimal route networks}
 \date{}
\author[1]{Kumar Sanat Ranjan}
\author[1]{Amit Ashok Pawar}
\author[1]{Arnab Roy}
\author[1]{Sandeep Saha}

\affil[1]{Department of Aerospace Engineering, Indian Institute of Technology Kharagpur, West Bengal, 721302, India}

\begin{document}
 
 \maketitle
\pagenumbering{arabic}
\begin{justify}
\begin{abstract}
The intriguing annual migration of the dragonfly species, \textit{Pantala flavescens} was reported almost a century ago \citep{fraser1924survey}. The multi-generational, transoceanic migration circuit spanning from India to Africa is an astonishing feat for an inches-long insect. Wind, precipitation, fuel, breeding, and life cycle affect the migration, yet understanding of their collective role in the migration remains elusive. We identify the transoceanic migration route by imposing a time constraint emerging from energetics on Djikstra's path-planning algorithm. Energetics calculations reveal a \textit{Pantala flavescens} can endure 90 hours of steady flight at 4.5m/s. We incorporate active wind compensation in Djikstra's algorithm to compute the migration route from years 2002 to 2007. The prevailing winds play a pivotal role; a direct crossing of the Indian Ocean from Africa to India is feasible with the Somali Jet, whereas the return requires stopovers in Maldives and Seychelles. The migration timing, identified using monthly-successful trajectories, life cycle, and precipitation data, corroborates reported observations. Finally, our timely sighting in Cherrapunji, India (25.2N 91.7E) and a branched network hypothesis connect the likely origin of the migration in North-Eastern India with \textit{Pantala flavescens}'s arrival in South-Eastern India with the retreating monsoons; a clue to their extensive global dispersal.
\end{abstract} \hspace{10pt}

\section{Introduction}
Every year millions of insects fly thousands of kilometers for conducive breeding habitats and foraging grounds \cite{dingle2014migration,holland2006and,chapman2015long}. The sheer volume of the migration has tremendous consequences on the global ecology \cite{chapman2015long,bauer2014migratory}. Migratory insects connect distant ecosystems across oceans, transport nutrients and propagules over long distances, and structure food webs  \cite{bauer2014migratory}.
Although the consequences on the food cycle, distribution of nutrients, pollination, modulating disease-causing micro-organisms \cite{satterfield2020seasonal}, panmixia \citep{troast2016global} are of universal significance, predicting transoceanic migration route for insects has had limited success \citep{hobson2021long,hedlund2021unraveling}. Identifying the migration route accurately could help understand gene flow and population dynamics of pests \citep{cao2019genetic}, disease outbreaks due to parasites \citep{huestis2019windborne} and locate the migrant carcass which drives seasonal transport of nutrients like nitrogen and phosphorus \cite{satterfield2020seasonal}. Improving the prediction of the migration route and timing is a prerequisite to applying the concept of `migratory connectivity' to insects \citep{gao2020migratory}, assisting population balance and pest control strategies and aiding annual cycle studies \cite{marra2015call}. Furthermore, studies show the population of the migrating insects declining \cite{dirzo2014defaunation,sanchez2019worldwide,hallmann2017more}. An increase in surface temperature shifts the overwintering sites away, thus stretching the migration range \citep{zeng2020global}. Consequently, long-range migration becomes more strenuous, leading to ecological concerns regarding the arrival of migrant species that regulate the local vector population and outbreak of diseases \citep{satterfield2020seasonal, huestis2019windborne}. Therefore, assessing the impact of climate change on the survivability of the migration relies intrinsically on route identification. 
 \section{Results}
 \subsection{Transoceanic migration} 
The annual migration circuit of dragonfly species, \textit{Pantala flavescens}, is a multi-generational, transoceanic path \cite{anderson2009dragonflies} spanning $14000-18000$ km from India to Africa \cite{Hobson2012}. Intriguingly, \textit{P. flavescens} crosses the Indian ocean from Africa to India without stopovers with assistance from winds associated with the Inter-Tropical Convergence Zone (ITCZ); an extraordinary feat for an insect with wings a few inches wide \citep{Anderson2009,hedlund2021unraveling}. \citet{anderson2009dragonflies} proposed the route India-Maldives-Seychelles-Africa-India (Fig. \ref{fig:PF_route}) based on his observations. Migratory routes range from simple round trips to complex circuits with merging and splitting branches \citep{satterfield2020seasonal} and the entirety of the transoceanic migration network of \textit{P. flavescens} is a subject of multiple contemporary investigations \citep{borisov2020seasonal,hobson2021long,hedlund2021unraveling}. Our understanding of transoceanic migration of \textit{P. flavescens} has been enhanced by wind trajectory analysis \citep{hu2021environmental,hedlund2021unraveling} that reveals how wind assists migration and stable isotope analysis \citep{Hobson2012} which indicates probable origins of the migration circuit. Yet, there remain crucial but inexplicable observations  \citep{hedlund2021unraveling}: For instance, the arrival of \textit{P. flavescens} at Maldives \citep{Anderson2009} and Seychelles \citep{bowler2003odonata} which are sparsely situated islands on the migratory route during a specific period of the year. Similarly, \citet{Hobson2012} hypothesized that the migration originates in North and East India, while \citet{Anderson2009} noted the arrival of \textit{P. flavescens} in South-East India and Sri Lanka with retreating monsoons. 
\begin{figure}[htbp]
    \centering
    \includegraphics[width=0.7\textwidth]{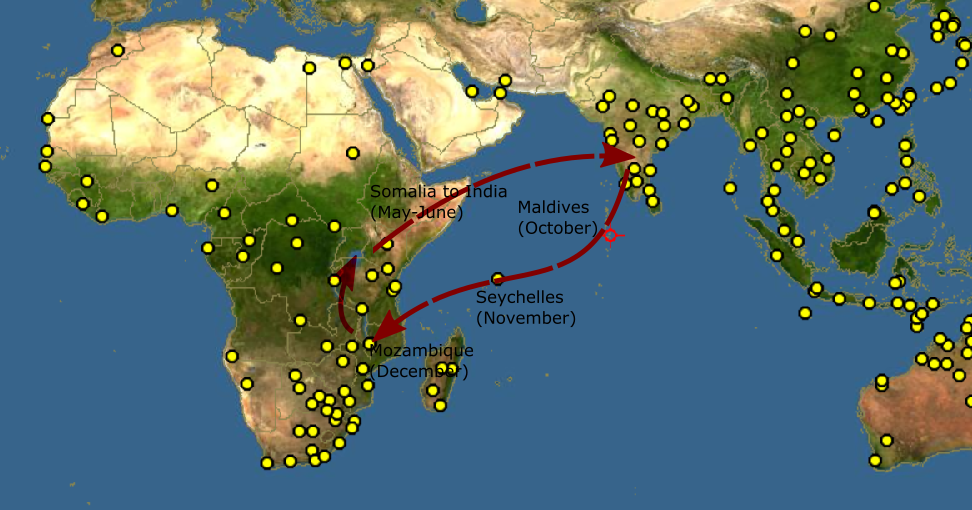}
    \caption{Proposed migration route of \textit{Pantala flavescens} \citep{Anderson2009}. The yellow dots indicate locations where \textit{Pantala flavescens} has been reported, Source: \href{http://www.discoverlife.org}{http://www.discoverlife.org}.}
    \label{fig:PF_route}
\end{figure}
Field investigations are challenging in the absence of sophisticated lightweight trackers that can be employed over a large geographical extent for studying the migratory patterns of \textit{P. flavescens} \citep{drake1995insect,holland2006and,hobson2021long,hedlund2021unraveling}. Existing migration studies \citep{chapman2010flight,wu2018advanced,westbrook2016modeling,chapman2015detection} successfully explain important features like heading and displacement; yet understanding transoceanic migration involves additional challenges. Predicting the migration timing of \textit{P. flavescens} that is consistent with the precipitation, wind patterns at the reported altitude, and the sightings over the entire migration circuit is daunting for multiple reasons. A self-consistent approach must account for the energetics \citep{hedlund2021unraveling} of the migrant for estimating the endurance while predicting the wind-assisted trajectory for destinations situated on islands. The trajectory analyses by \citet{hobson2021long,hedlund2021unraveling} assume that \textit{P. flavescens} fly downwind; remaining on course during ocean crossings would require wind compensation, alongside wind assistance, to reach stopovers. We propose a complementary approach using an optimal path-planning algorithm \citep{dijkstra1959note} applied to micro-aerial vehicles (MAVs) that allows wind compensation. The constraints for MAVs and \textit{P. flavescens}, indeed, are analogous. The migrants must choose an altitude with favorable tailwind while suitably compensating for wind direction \citep{srygley2003wind} and fly on an intended path like MAVs. Further,  the stopovers can only be on an \emph{island} similar to waypoints in MAV missions. In addition, the stopovers must be located such that the migrant can reach the destination with limited energy reserves like MAVs. The combination of all these constraints makes the timing of migration the key to deciding the success of the migration. The optimal route does not constitute the migration trajectory \emph{per se}; rather, migration occurs along a near-optimal path, determining which using trackers is nigh impossible. We present the optimal trajectory and verify the consistency of the timing for the entire migration with the reported observations. We adapt an energetics model \citep{pennycuick2008modelling} for \textit{P. flavescens} and find that the utmost duration can fly at the optimal velocity,  $4.5m/s$, for maximizing range is approximately $90 h$ ( see sec. 5.2); an estimate consistent with \citet{hedlund2021unraveling,hobson2021long}. Thereafter, we apply \textit{Dijkstra's} algorithm \citep{dijkstra1959note} to obtain the optimal path for a minimum time under the influence of wind for the transoceanic legs of the migration. An optimal trajectory requiring a flight time less than $90 h$ is considered successful. The migration time window is identified from the number of monthly-successful trajectories over a period of six years (2002-2007) \citep{Anderson2009} and compared to the local precipitation data and sightings of \textit{P. flavescens} for each stopover.
\subsection{Optimal Path}
Fig. \ref{fig:Optimal path} shows the optimal path for the transoceanic legs of the \textit{P. flavescens}'s migration circuit and corroborates existing observations \citep{fraser1924survey, Corbet1962, Anderson2009, Hobson2012, hedlund2021unraveling}. The flight from Somalia (Location 5) to India (Location 1) is the longest leg ($2802 kms$) but requires merely $45 h$ and no stopovers because of strong tailwinds ($\sim 15m/s$). The migration from India (Location 1) to Mozambique (Location 4) is more precarious and involves stopovers at Locations 2 (the Maldives) and 3 (Seychelles). The trajectory from India to Maldives is tortuous and requires $48 h$ for $1137 kms$ only. The stark difference between the time required and the distance covered in the two optimal paths highlights the pivotal role of wind. Further, comparing the optimal path with the trajectory of a passive tracer reveals that active wind compensation is imperative for \textit{P. flavescens} to cross the Indian ocean; a passive tracer convecting downwind strays from the optimal paths (see Fig. \ref{fig:Optimal path}(c-e)). The flight from Maldives to Seychelles ($2565 kms$) and the subsequent flight from Seychelles to Mozambique ($2056 kms$) are even more critical in the success of the migration because the time required is $90 h$ and $81 h$ respectively, nearly extinguishing their energy reserves. Therefore, we deduce that  \textit{P. flavescens} actively perform wind compensation \citep{srygley2003wind,chapman2010flight} while flying at a minimum pace to  maximize their range \citep{hedlund2021unraveling}.
\begin{figure}[h!]
\begin{center}
\subfloat[]{\includegraphics[width=0.9\textwidth]{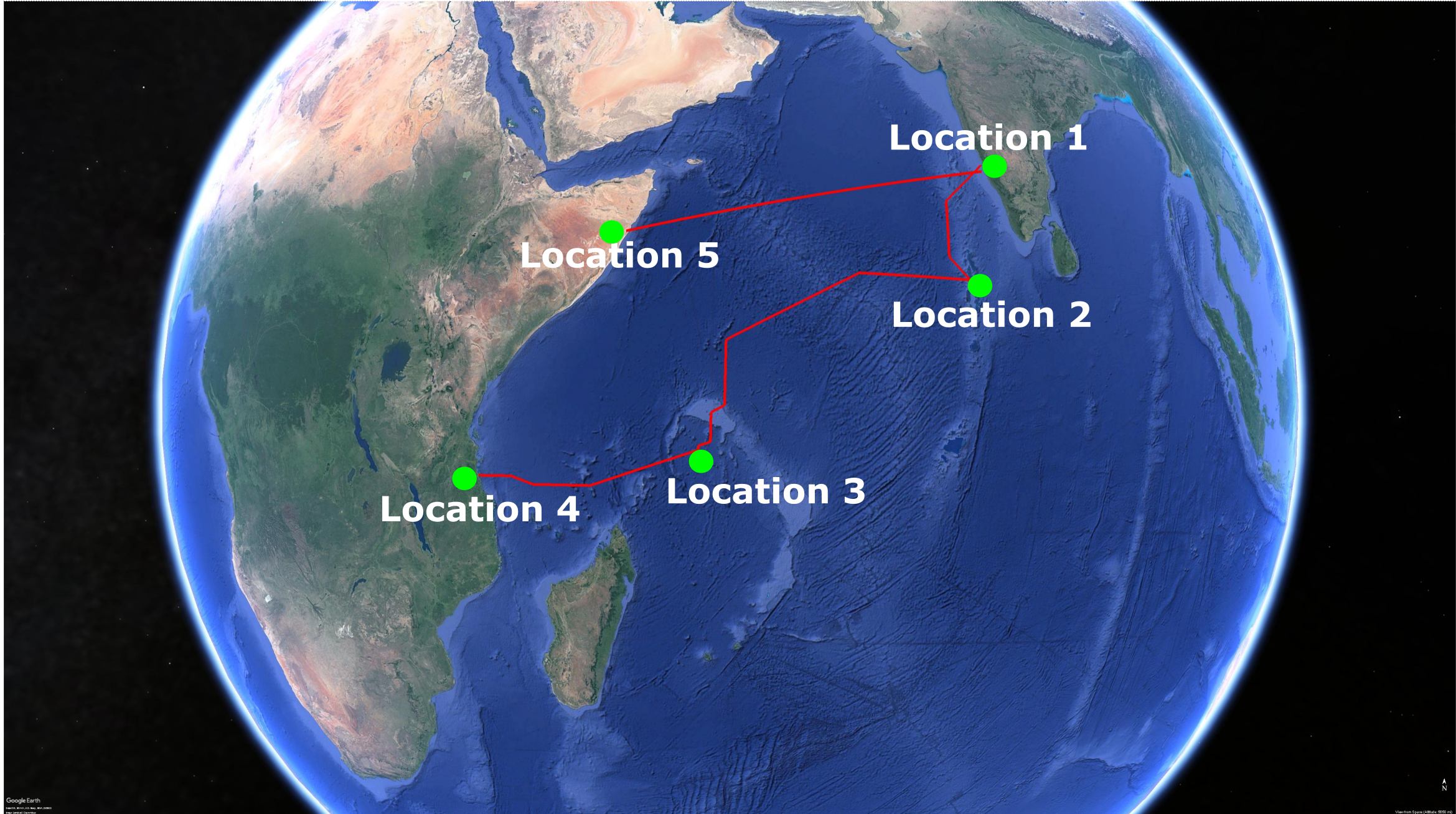}}\\
\subfloat[]{\includegraphics[width=0.45\textwidth]{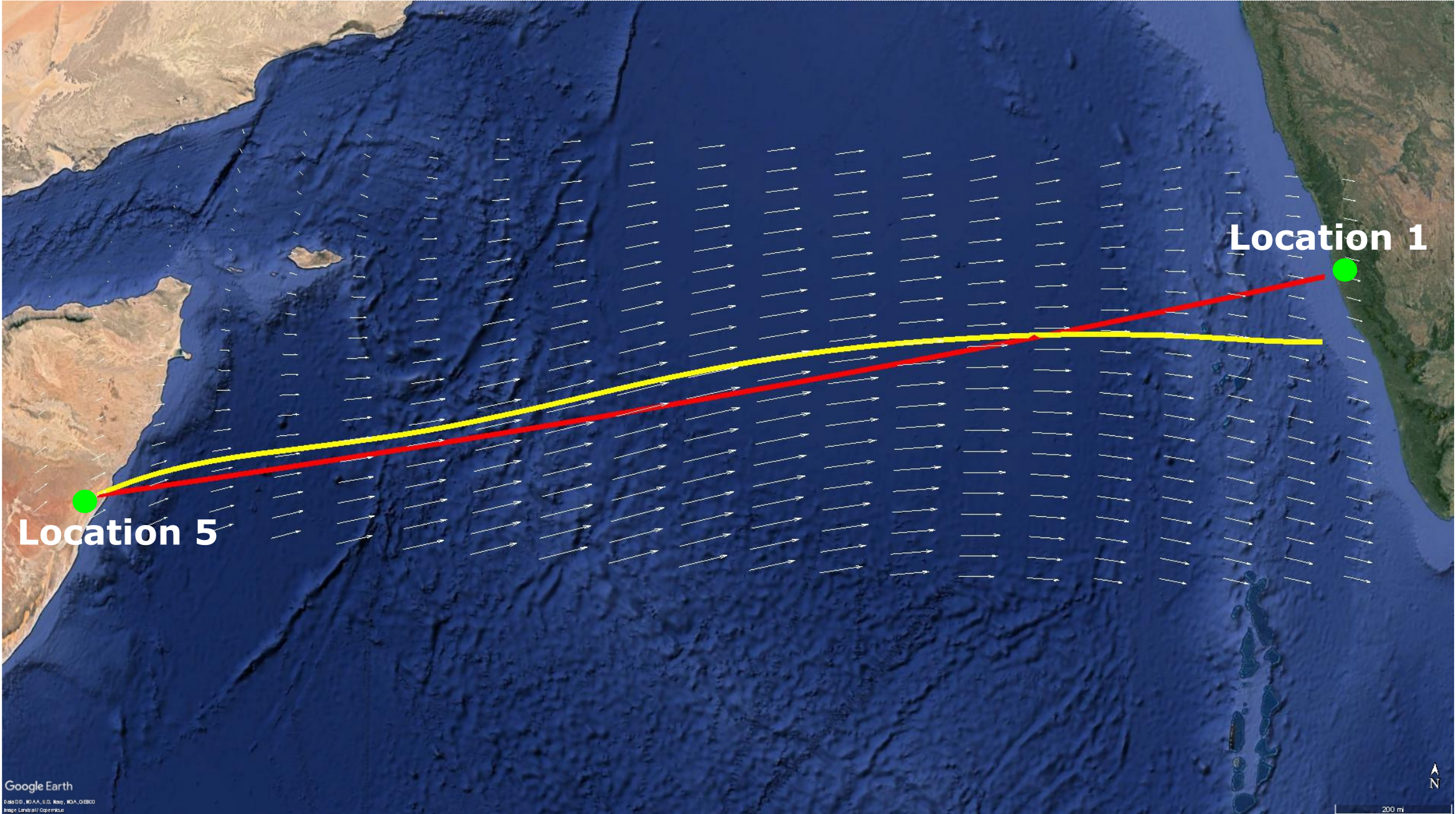}}\quad
\subfloat[]{\includegraphics[width=0.45\textwidth]{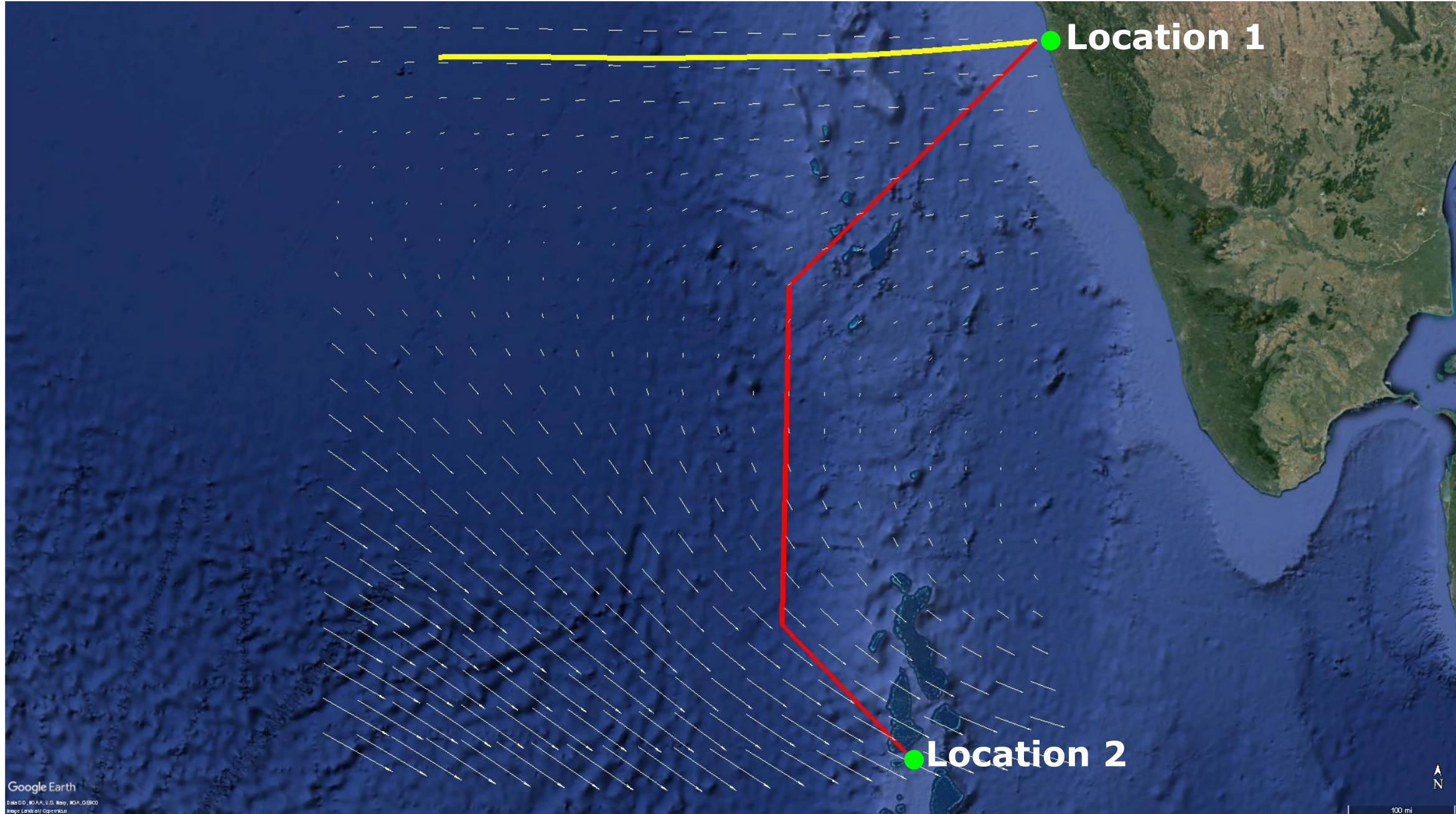}}\\
\subfloat[]{\includegraphics[width=0.45\textwidth]{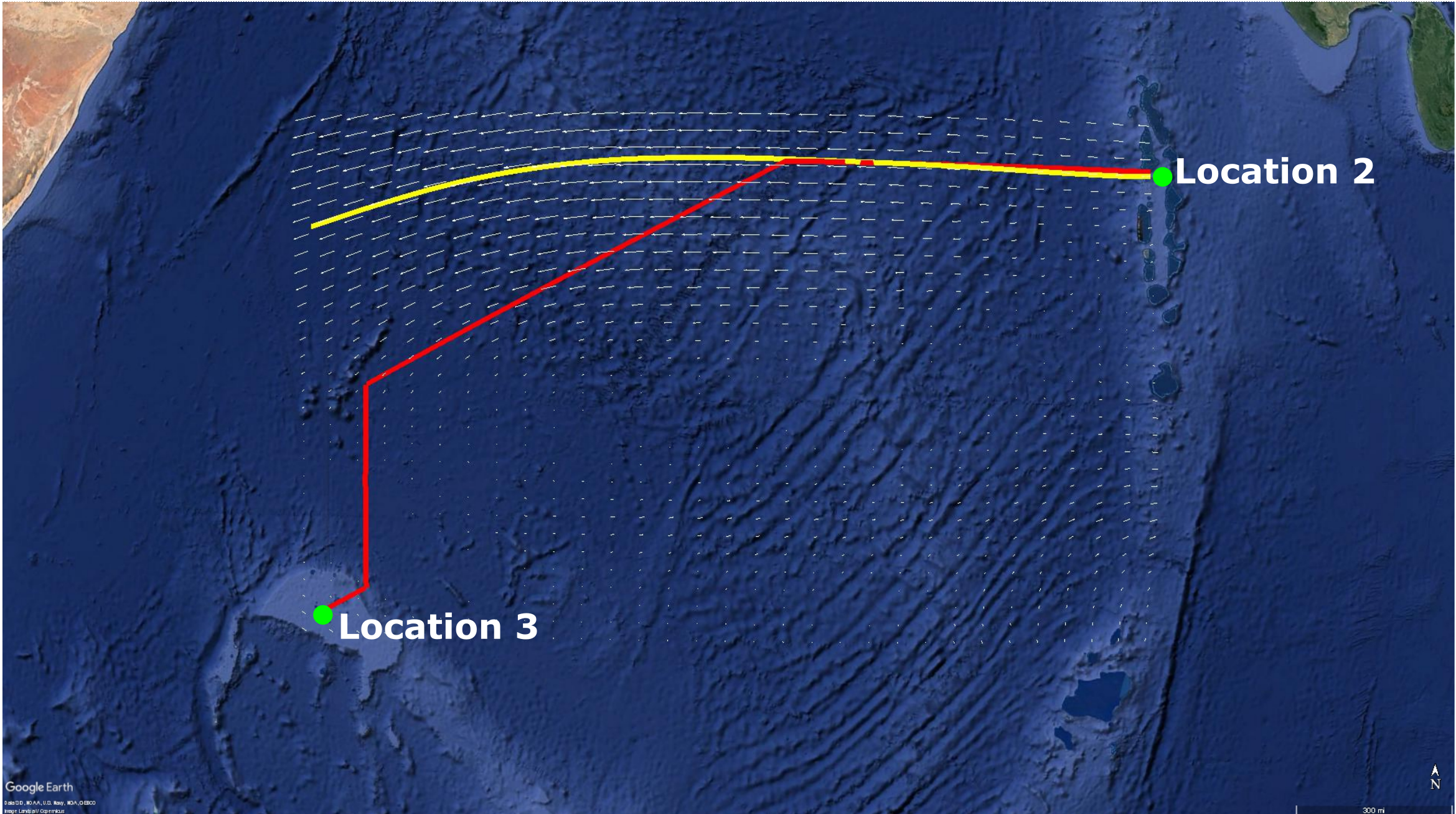}}\quad
\subfloat[]{\includegraphics[width=0.45\textwidth]{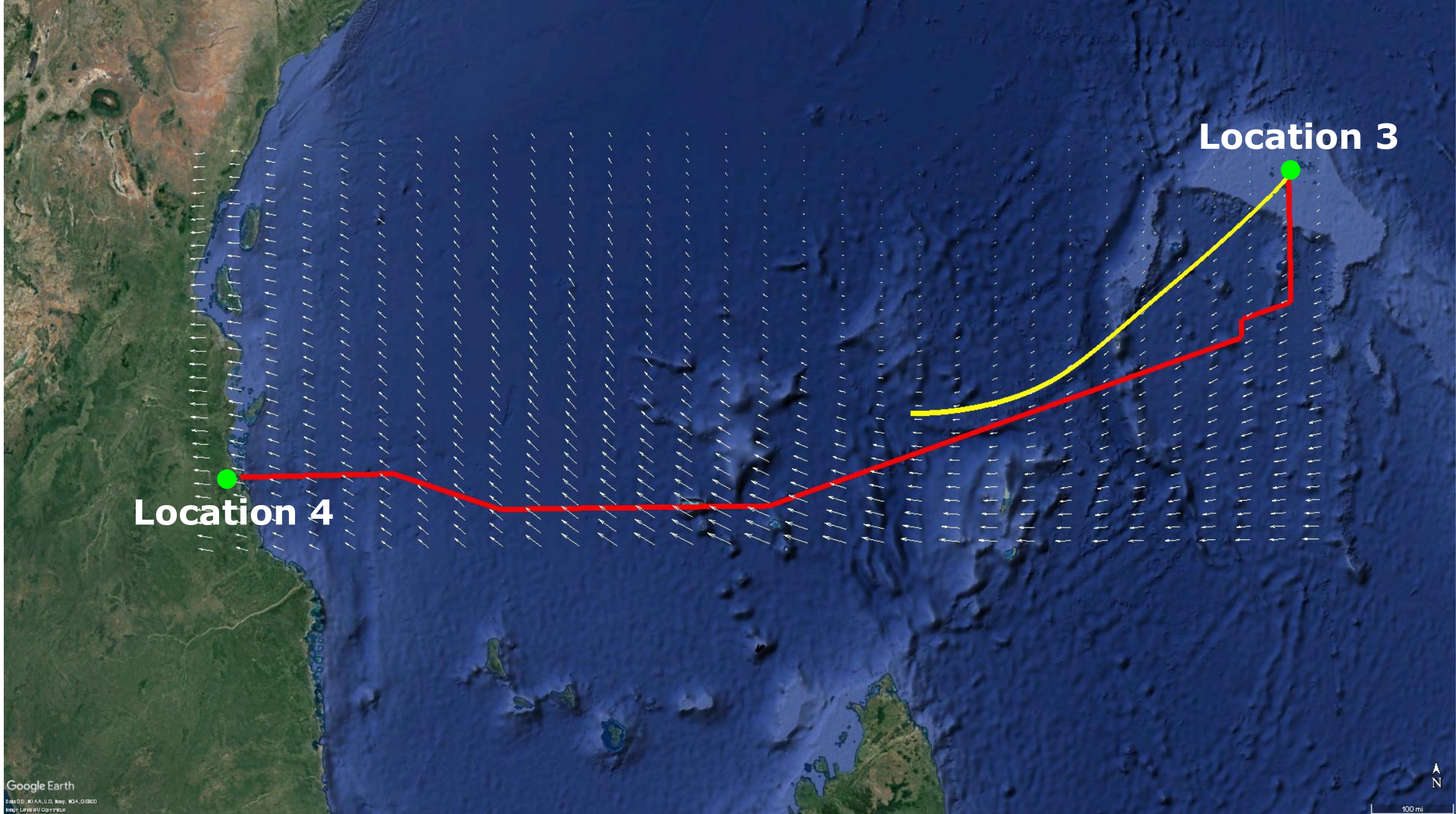}}\\
\end{center}
\caption{%
a) Optimal path (\textcolor{red}{\rule{0.5cm}{1mm}}) and the trajectory of a passive tracer (\textcolor{yellow}{\rule{0.5cm}{1mm}}) transported by the local wind (shown as vector field) at 850 hPa for the transoceanic legs of \textit{P. flavescens}'s annual migration circuit. b) Location 5 [7.5N	49.5E](Somalia) to Location 1 [13.5N	74.5E](India) on 15/06/2016: 2802 kms in 45 h; c) Location 1 to Location 2 [5N	73E](Maldives) on 15/10/2016: 1137 kms in 48 h; d) Location 2 to Location 3 [4.5S	55.5E](Seychelles) on 17/11/2016: 2565 kms in 90 h; e) Location 3 to Location 4 [10S	39.5E](Mozambique) on 18/12/2016: 2056 kms in 81 h.}%
\label{fig:Optimal path}
\end{figure}
\subsection{Migration time window}
\textit{P. flavescens} are obligate migrants, whose migration timing shadows the movement of the Inter-Tropical Convergence Zone, pursuing evanescent pools for breeding \citep{Anderson2009}. Indeed, their sightings at the stopovers reported in the literature (see SI Table 6) and the months of high precipitation (see Fig. \ref{fig:Bargraph}) approximately coincide. Therefore, migration occurs \emph{only when} both wind assistance and rainfall are available simultaneously, creating the migration time window. We present the total monthly successful trajectories for years 2002-2007 in Fig. \ref{fig:Bargraph} to identify the migration time window. 
Fig. \ref{fig:Bargraph}(a) reveals that precipitation in India (Location 1) peaks between May and September and also the successful trajectories from Somalia to India, thus constituting the time window favourable for migration. The arrival of \textit{P. flavescens} in June-July, aided by the Somali Jet, is well documented \citep{corbet2004dragonflies,Anderson2009,may2013critical}. Thereafter, the migrants breed and the offsprings typically emerge after 45-60 days \citep{corbet2004dragonflies,suhling2004field,kumar1984life}. Furthermore, Fig. \ref{fig:Bargraph}(b) reveals that the wind assistance from India to Maldives is available except from June to August. The ITCZ passes over southern India and Maldives in October, bringing rainfall, and the onset of migration from India to Maldives lasting till December. Upon arrival in Maldives the migrants either breed \citep{olsvik1992dragonfly} in evanescent pools created by rainfall \citep{corbet2004dragonflies} or migrate towards Seychelles \citep{campion1913no,bowler2003odonata,Anderson2009} to breed \citep{wain1999odonata}. Breeding in Maldives might be preferable as crossing the Indian Ocean to reach Seychelles between October and December, although not impossible, remains unfavourable (see Fig. \ref{fig:Bargraph}(c)). Thereafter the precipitation in and migration to Seychelles is more favourable from January to March, corroborating sightings \citep{samways1998establishment,samways2010tropical,Alphonse,SBRC}. The onward journey to the African mainland is also favoured by the wind and precipitation in the period December to March, as seen in Fig. \ref{fig:Bargraph}(d). Indeed, sightings from Mozambique and Malawi, South Africa, and Lake Tanganyika support our findings (see Table 6). \textit{Pantala flavescens} may breed twice on the African mainland before migrating to India, thus spawning 4-5 generations every year \citep{corbet2004dragonflies}.
\begin{figure}[h!]\centering
\subfloat[]{\includegraphics[width=0.45\textwidth]{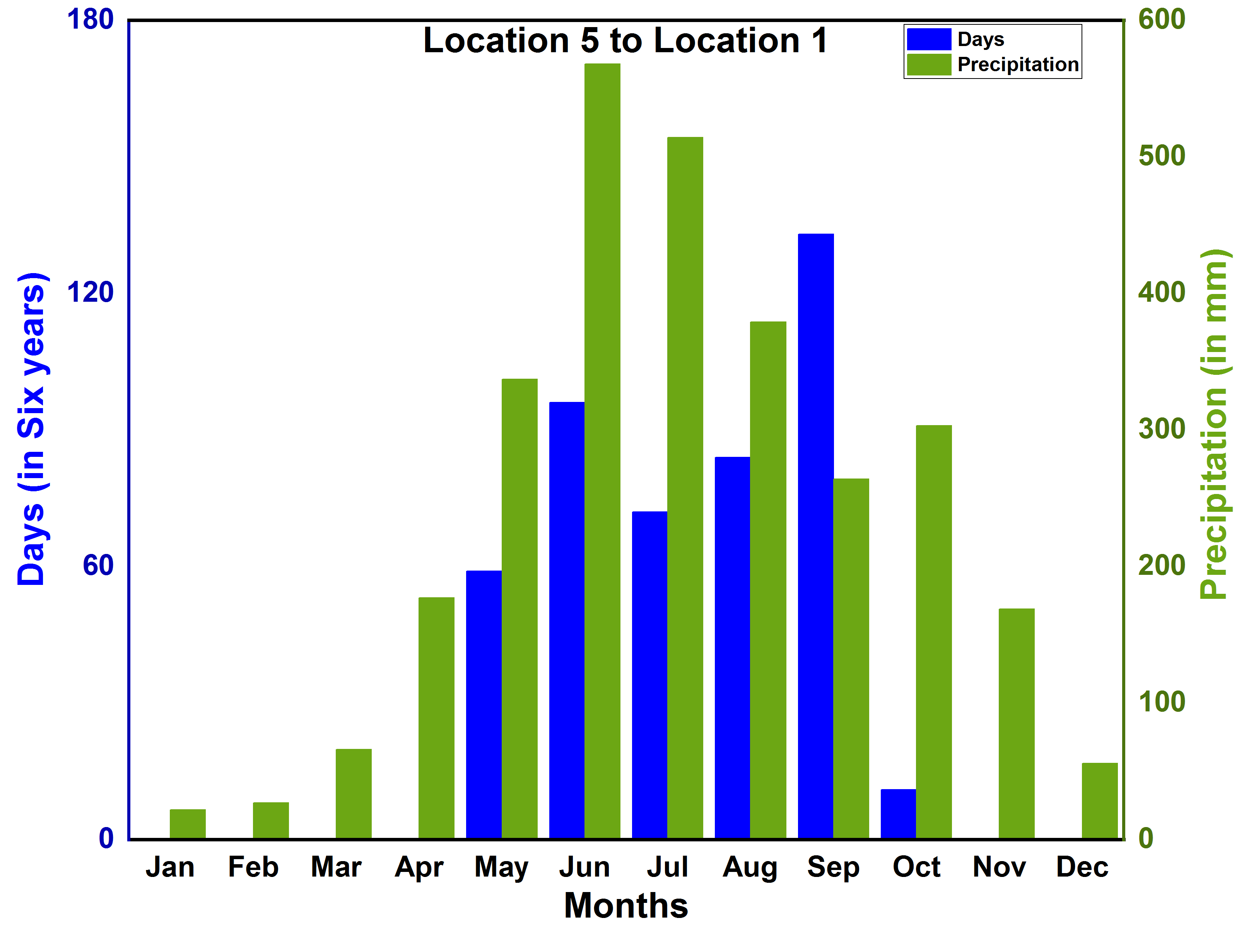}}\quad
\subfloat[]{\includegraphics[width=0.45\textwidth]{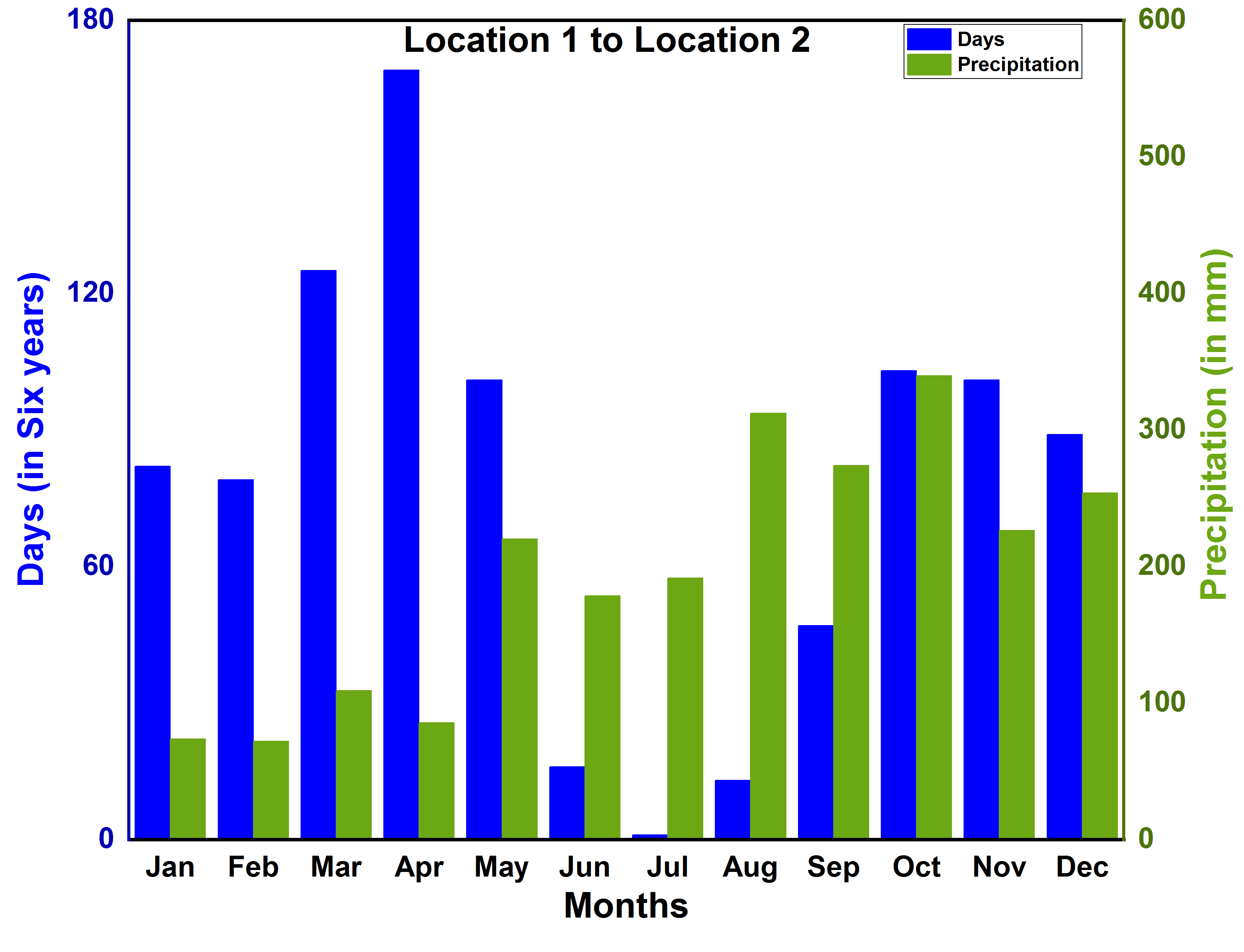}}\\
\subfloat[]{\includegraphics[width=0.45\textwidth]{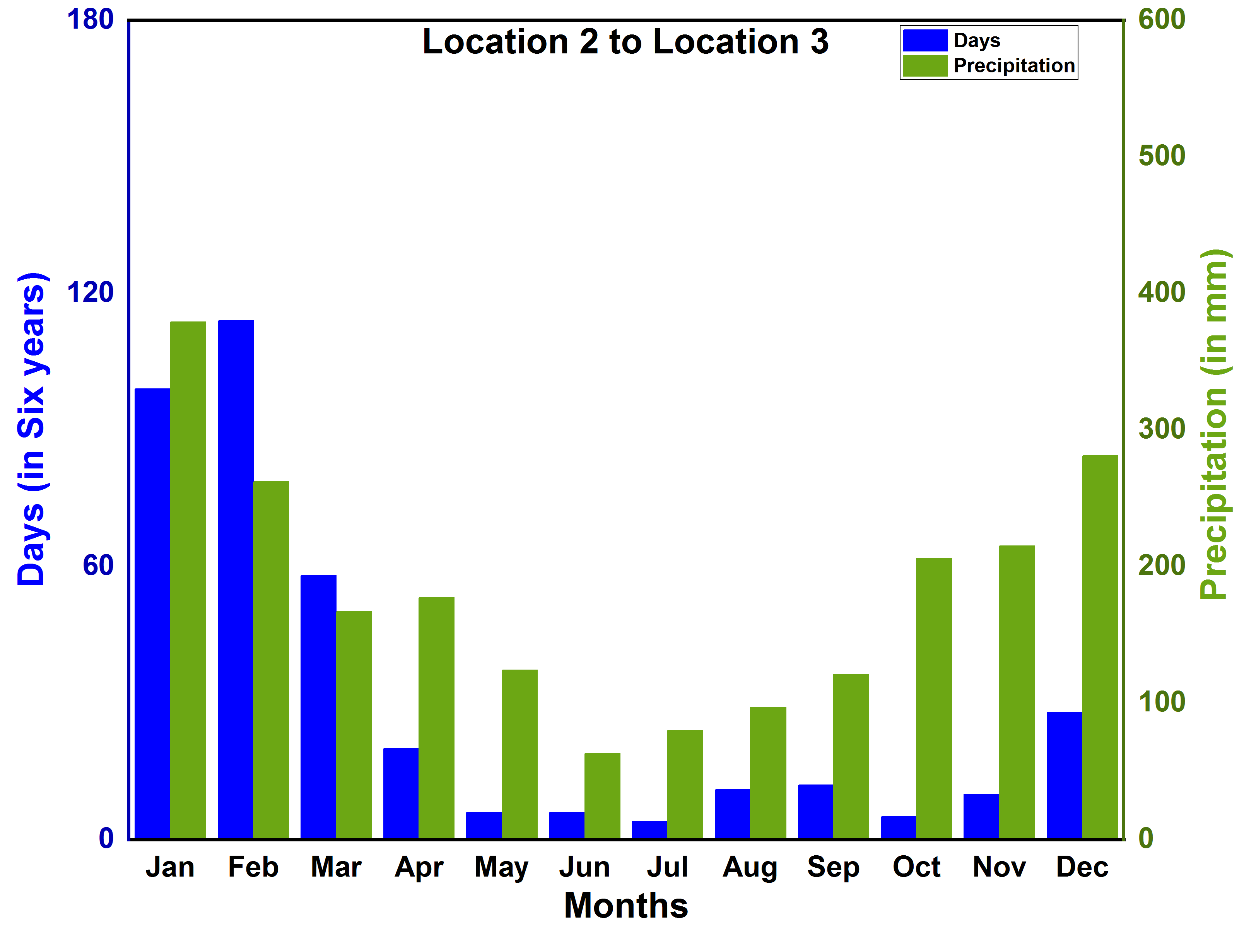}}\quad
\subfloat[]{\includegraphics[width=0.45\textwidth]{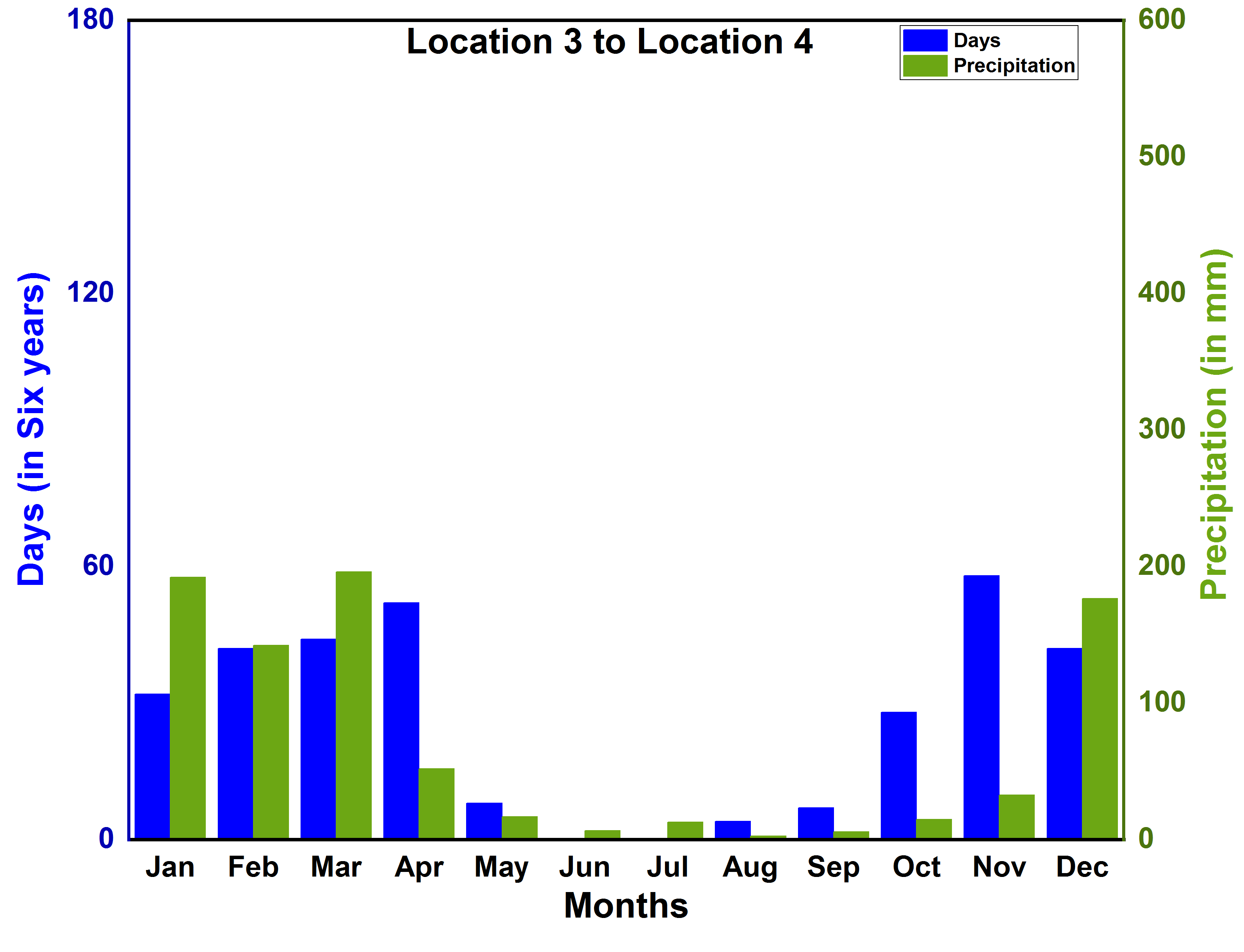}}
\caption{%
 Month-wise distribution of the total number of days out of six consecutive years (2002-2007) on which the migration is completed within 90 hours for all four legs. Monthly precipitation at a) Location 1 b) Location 2, c) Location 3, d) Location 4}%
\label{fig:Bargraph}
\end{figure}
\subsection{Alternate routes, branching and implications on dispersal} 
Existing studies \citep{Anderson2009,borisov2020seasonal,hedlund2021unraveling} suggest an alternate route for \textit{P. flavescens} to cross the Indian Ocean directly from various departure sites in India and the Maldives to arrive in Somalia during  November and December. We computed the successful trajectories for emigration from Locations 1 (India), 2 (Maldives) to Location 5 (Somalia), and indeed winds are not favourable until September (see Fig. \ref{fig:branching}(a)), and most successful trajectories occur in December, consistent with \citet{hedlund2021unraveling}. The passage of the ITCZ in October through this region renders the direct crossing feasible. November seems more favourable for migration than December because there is more precipitation in Somalia.
\begin{figure}[h!]\centering
\subfloat[]{\includegraphics[width=0.5\textwidth]{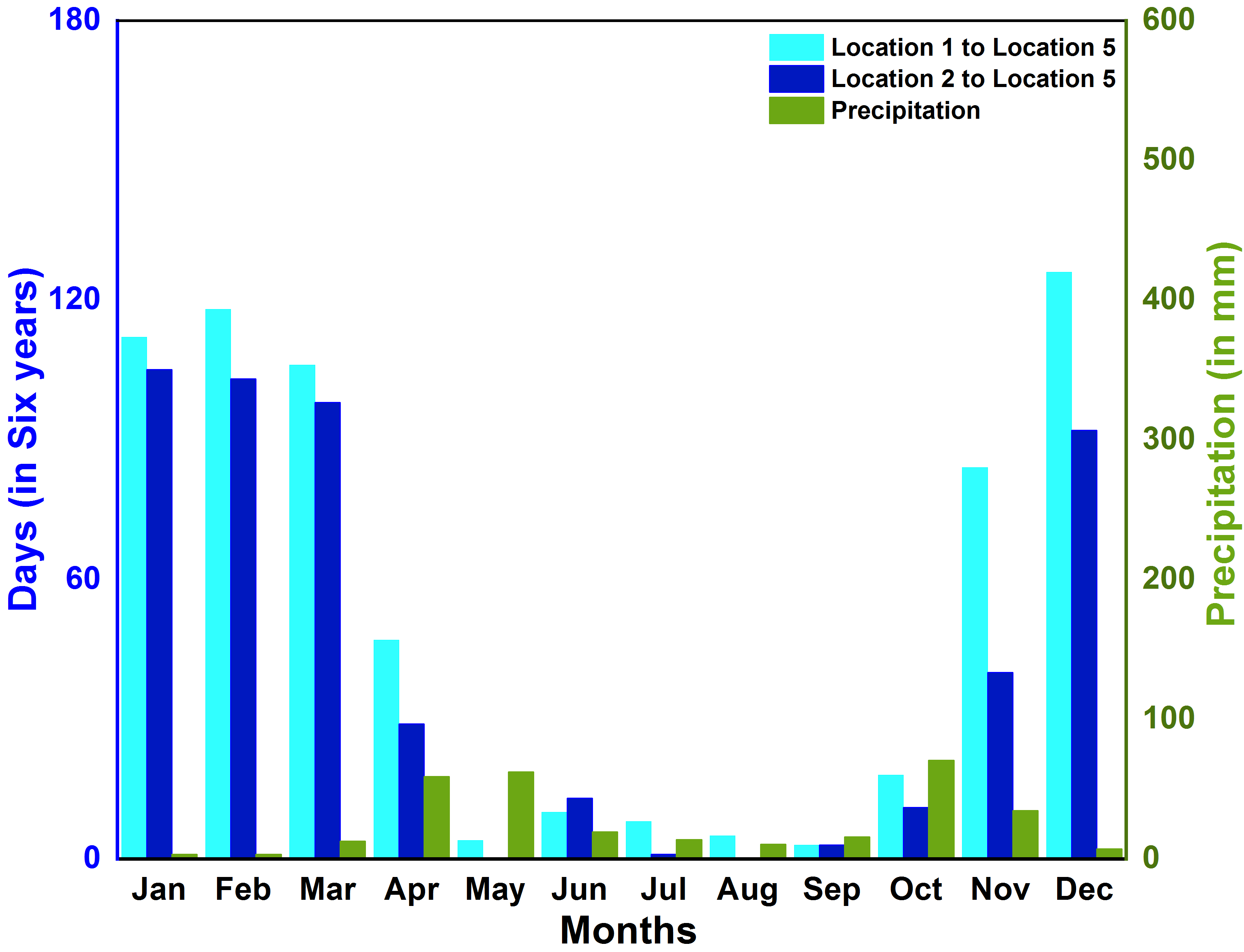}}\quad
\subfloat[]{\includegraphics[width=0.4\textwidth]{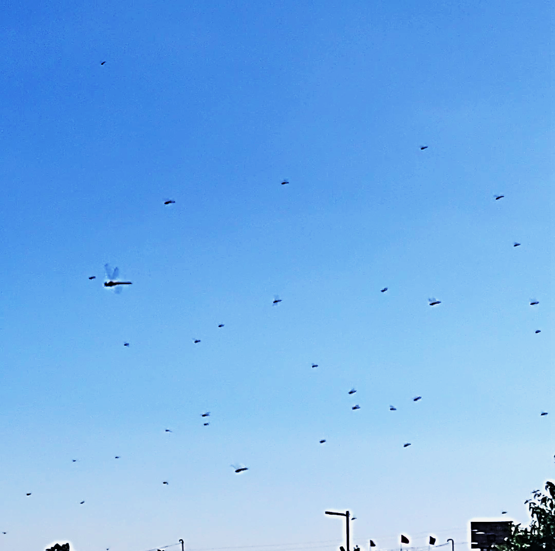}}\\
\subfloat[]{\includegraphics[width=0.9\textwidth]{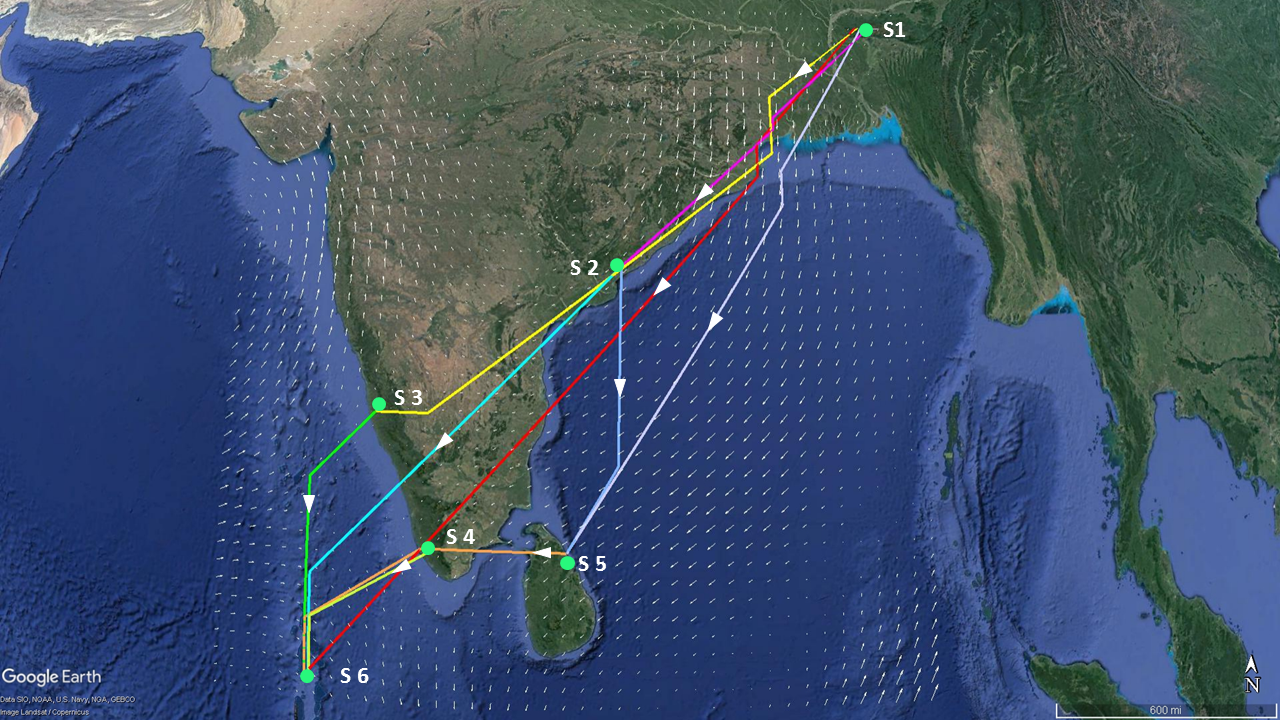}}
\caption{(a) Month-wise distribution of the total number of days out of six consecutive years(2002-2007) on which the migration is completed within 90 hours for two alternate routes; Location 1 to Location 5 and Location 2 to Location 5, and monthly precipitation at Location 5. (b) Sighting of migration swarm of \textit{P. flavescens} at Cherrapunji, Meghalaya(S1) on $2^{nd}$ November 2019. (c) Branching network of various possibilities of reaching Maldives(S6) starting from Cherrapunji(S1) on $2^{nd}$ November 2019, with stopovers at Visakhapatnam [17.5N 82.5E](S2), Mangalore[13N 75E] (S3), Thiruvananthapuram[9N 77E] (S4) and in Sri Lanka[9N 81E] (S5).}
 \label{fig:branching}
\end{figure}
The possibility of the alternate route implies the existence of branching networks \citep{drake1995insect,satterfield2020seasonal}. The concept of a branched network allows a complex migratory pattern to emerge naturally and connect unexplained albeit important observations reported hitherto. For instance, \textit{P. flavescens} arrive in south-eastern India and eastern Sri-Lanka along with the NE monsoons from October to December \citep{corbet1988current,Anderson2009}. Furthermore, the origin of \textit{P. flavescens} reaching Maldives is speculated as north-eastern India \citep{Hobson2012,hedlund2021unraveling}. We sighted \textit{P. flavescens} swarms in Cherrapunji (25.2N 91.7E, NE India (S1)) on $1^{st}$ and $2^{nd}$ November 2019 (see fig. 
\ref{fig:branching}(b)). The swarm exhibited a well-coordinated motion predominantly from the north-east to the south-west direction (see supplementary movie) aligned with the local wind on those days, indicating a destination potentially in south-eastern India or Sri Lanka. A branched migration network coupled with the sighting in Cherrapunji allows us to conjecture that, indeed, the transoceanic migration of \textit{P. flavescens} plausibly originates in NE India with stopovers in SE India and Sri Lanka. We investigate the branched networks on this pathway further using optimal paths (see fig. \ref{fig:branching}(c)). A direct flight from Cherrapunji to Maldives requires the migrant to cover a distance of around $3000km$, requiring approximately $115 h$, which is significantly higher than the threshold of $90h$. There is ample land mass between these two sites, and multiple refuelling stopovers may be anticipated. Fig. \ref{fig:branching}(c) shows the branching network of various possibilities of reaching S6 from S1 with stopovers at Visakhapatnam (S2), Mangalore (S3), Thiruvananthapuram (S4) and in Sri Lanka (S5)(see SI Table 7). From the figure, it can be observed that S2 and S4 are potential stopovers as they are in the path of multiple optimal routes, and they are more likely to be selected by migrating \textit{P. flavescens}. The timely sightings at Cherrapunji, SE India and Sri Lanka and the branched networks revealed by the optimal paths lend further credibility to the NE India being the origin of the transoceanic migration of the \textit{P. flavescens}.

The branched network in Fig. \ref{fig:branching}(c) provides a glimpse of the complex migratory network of P. flavescens, potentially spanning Asia and Africa. The appearance of P. flavescens in Japan, China, Indonesia, Sri Lanka, NE India, and southern Islands of the Indian ocean such as Amsterdam Island and Chagos Archipelago(see SI Table 6) leads to speculation that there is a possibility of branching and dispersal of migrating P. flavescens from all the locations that are part of a more complex migratory circuit spanning Asia, Africa and beyond. This migration significantly impacts global ecology, and its success is linked to any stressors of the climate and local ecological systems. There have been reports of disappearing islands \citep{farbotko2010global} which are detrimental to the migration of \textit{P. flavescens} and, in turn, the global ecology.
\section{Conclusions} 
The migration from India to Africa starts from October with stopovers in Maldives and Seychelles, as suggested by \citet{Anderson2009}. The arrival timing predictions for Maldives, Seychelles, and Mozambique are in agreement with prior observations \citep{Anderson2009,olsvik1992dragonfly,wain1999odonata,campion1913no}(see SI Table 6). The migration time window indicates the possibility of breeding in Maldives and Seychelles, which is consistent with \citet{olsvik1992dragonfly}'s and \citet{wain1999odonata}'s observations, respectively. The direct crossing of the Indian ocean aided by the Somali Jet is feasible, May onwards, during the return migration from Africa to India, corroborating  previous results \citep{Anderson2009,corbet2004dragonflies,hedlund2021unraveling}. An alternate route involving the direct crossing of the Indian ocean for the onward journey from India and Maldives to the African continent is also feasible in November and December, as reported by \citet{hedlund2021unraveling}. Furthermore, the identification of alternate routes implies the existence of a branched and complex network of migratory routes. We also correlated our spotting of swarms of migrant \textit{P. Flavescens} in Cherrapunji, India on $1^{st}, 2^{nd}$ November 2019 with two seemingly disconnected results in the literature; \citet{Hobson2012,hedlund2021unraveling} predicted that the migration of \textit{P. Flavescens} begins in NE India and \citet{corbet1988current,Anderson2009} reported the arrival of \textit{P. Flavescens} in SE India and Sri Lanka. Indeed all the observations fit into an extensive, branched migration circuit of \textit{P. Flavescens} that originates in NE India and has stopovers spread across the Indian subcontinent and might explain the closeness of the local populations \citep{christudhas2014genetic}. A branched migratory pattern with multiple stopovers also explained the widespread dispersal of \textit{P. Flavescens} throughout SE Asia and Africa(see SI Table 6 for Ref.). The \textit{P. Flavescens} initiate migration from various sites and probably halt at stopovers when they find cues for favorable conditions. Furthermore, the availability of micro-insects \citep{Anderson2009} and aerial planktons \citep{may2013critical,troast2016global} for feeding en route could increase the continuous flight duration and, consequently, the range. The migration in swarms is also likely to reduce the aerodynamic drag on each individual. Thus the energetics constraints imposed in our model are conservative and underestimate the probability of migration success. Our conclusions lead to new questions like the impact of climate change on and the role of the swarm dynamics in insect migration. 
\end{justify}

\begin{justify}
\section{Materials and Methods}
\subsection{Dragonfly Energetics Model (DEM)}
We designate the energetics model applied to \textit{P. Flavescens} as the Dragonfly Energetics Model (DEM). Dragonfly species \textit{P. flavescens} cover large distances in a single flight during the trans-oceanic migration. The energetics involved in their migration though important has not yet been understood. However, there have been significant advances in bird migration theories, which have led to accurate calculations of energetics. In the present study, we develop a computational model for \textit{P. flavescens}'s migration energetics by adapting the model proposed by \citet{pennycuick2008modelling}. The model is motivated by the long range transport aircrafts that involve energetics that are similar to the migration process, and is generic to any flying species, and has been applied to insects \citep{warfvinge2017power} and birds \citep{pennycuick2008modelling}.

\textbf{Energetics model for migration range, time of flight and maximum range velocity} 

\begin{figure}[h!]
    \centering
    \includegraphics[trim=70 50 10 20,clip,width=0.9\textwidth]{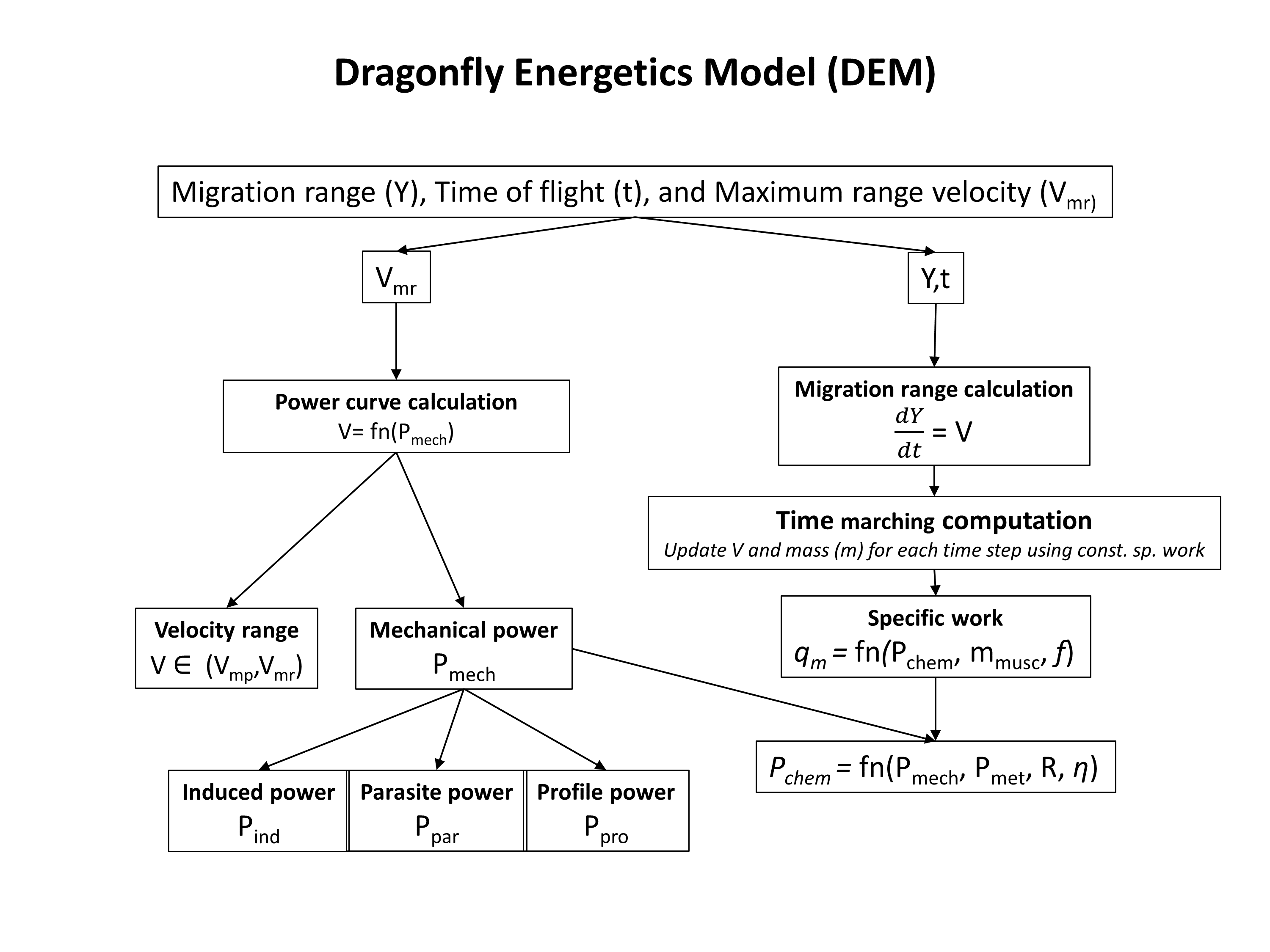}
    \caption{\textbf{Overview of migration energetics model} }
    \label{fig:DEM_flow_chart}
\end{figure}
The migration range and the time of flight are obtained by  numerically integrating the instantaneous migration speed. Fig. \ref{fig:DEM_flow_chart} shows an overview of the migration energetics model. The migration can last up to the point where the fat is completely burnt and defines the upper limit for the time of flight. The migration speed varies over time; for the major part of the course of migration, the migrant flies at a speed close to the maximum range speed ($V_{mr}$) that corresponds to the speed at which the lift to drag ratio is maximum. Before achieving $V_{mr}$ the migrant flies instantaneously at a speed that fulfills the ``constant specific work" (work done per unit mass) criterion. The criterion agrees with the observation that flight muscle fraction in migrants is nearly constant \citep{pennycuick2008modelling}. The constant specific work is obtained from the chemical power ($P_{chem}$), muscle mass ($m_{musc}$) and the wing beat frequency ($f$). The chemical power ($P_{chem}$) is obtained from the mechanical power ($P_{mech}$) required to fly and the metabolic power ($P_{met}$). 
Hence the energetics model involves two steps: Calculation of the power curve (the relationship between mechanical power and speed) and then migration range calculation using the power curve. We describe the associated parameters to compute the energetics model.


\subsubsection{Power curve calculation}
The required inputs for power curve calculation are three morphological parameters (mass($m$), wing span ($B$), and wing area ($S$)), gravity ($g$), and air density ($\rho$) at the desired height. The power curve is used to determine the total mechanical power ($P_{mech}$) required to maintain a horizontal flight, the minimum power speed ($V_{mp}$), and maximum range speed ($V_{mr}$). The minimum power speed ($V_{mp}$) is the speed at which the power required to fly is minimum. The maximum range speed ($V_{mr}$) can alternatively be viewed as the speed at which a flier can cover the maximum distance per unit of fuel consumed. The total mechanical power required to fly at any particular speed consists primarily of three components of power; the induced power ($P_{ind}$), the parasite power ($P_{par}$), and the profile power ($P_{pro}$).

\par
\textbf{Induced power}\\
The induced power is the rate at which the flight muscles of the insect have to provide work to impart downward momentum to the air at a rate that is sufficient to support the weight of the insect. The induced power is estimated using the actuator disc theory assuming a continuous beating of the wings as an actuator disc, and the pressure difference between the upper and the lower surface providing the aerodynamic force. The force multiplied by the induced power factor (k), which accounts for the loss in efficiency due to the real flapping of the wings, gives the real induced power (Eq. \ref{P_ind}).
\begin{equation}
P_{ind} = \frac{2k(mg)^2}{V_t{\pi}B^2\rho},    
\label{P_ind}
\end{equation}
where {$V_t$} is {true air speed}.\\

\textbf{Parasite power}\\
The parasite power is the rate of work required to overcome the drag acting on the insect's body, excluding the wings. The parasite power can be found from the drag acting on the body (Eq. \ref{P_par}),
\begin{equation}
P_{par} = \frac{{\rho}{V_t}^3S_bC_{Db}}{2}.
\label{P_par}
\end{equation}
where, {$S_b$} is the body frontal area, which is the maximum cross-sectional area of the insect,  and {$C_{Db}$} is the body drag coefficient.

\par
\textbf{Profile power}\\
A flying insect needs profile power to overcome the drag acting on the wings and it is essentially a consequence of the induced and parasite powers. The profile power is estimated from the minimum of the sum of induced power (Eq. \ref{P_ind}) and parasite power (Eq. \ref{P_par}) that is termed as the absolute minimum power ($P_{am}$) (Eq. \ref{P_am}); the power required to fly at the minimum power speed $V_{mp}$ (Eq. \ref{V_mp}). 
\begin{eqnarray}
P_{am} & = &\frac{1.05k^{3/4}m^{3/2}g^{3/2}S_b^{1/4}C_{Db}^{1/4}}{\rho^{1/2}B^{3/2}} \label{P_am} \\
V_{mp} & = & \frac{0.807k^{1/4}m^{1/2}g^{1/2}}{\rho^{1/2}B^{1/2}S_b^{1/4}C_{Db}^{1/4}} \label{V_mp}
\end{eqnarray}

The profile power is then set at a fraction ($X_1$) of the absolute minimum power $P_{am}$. Here $X_1 \equiv \frac{C_{pro}}{AR}$ 
and profile power, $P_{pro} \equiv X_1 P_{am}$.
The total mechanical power,$P_{mech}$, is then given by the summation of individual powers, 
\begin{equation}
P_{mech} \equiv P_{ind} + P_{par} + P_{pro}.    
\label{P_{mech}}
\end{equation}

In order to maximize the range of migration the maximum range speed, $V_{mr}$ also needs to be determined from the power curve . The speed $V_{mr}$ is obtained by drawing a tangent from the origin to the power curve \citep{pennycuick2008modelling}. The power curve ($P_{mech}$ versus velocity) is calculated over a range extending from the minimum power speed $V_{mp}$ (equation \ref{V_mp}) to the maximum range speed $V_{mr}$.


\subsubsection{Migration range calculation}
The migration range is obtained by solving the range equation $dY/dt = V$ where $Y$ is the distance covered from the source and $V$ is the instantaneous migration speed. The instantaneous speed, $V$, varies over time because the mass of the insect changes as a function of time due to burning of fat and protein; consequently the aerodynamic and morphological parameters evolve over time. The model incorporates these changes; the rate at which fuel burns depends on the chemical power ($P_{chem}$), the lift to drag ratio ($N$) and the wingbeat frequency ({$f$}). We initialize the migration range calculation with values obtained from the insect's measured morphological (for instance see SI table S4 ) and external parameters ($\rho$,g). \par
\textbf{Chemical power}\\
Chemical power is expended by an insect by burning fuel to generate the required mechanical power and support its metabolism. To determine chemical power $P_{chem}$ (Eq. \ref{P_chem}); the mechanical power ($P_{mech}$), basal metabolic rate ($P_{bmr}$), conversion efficiency ($\eta$), and respiration ratio ($R$) are incorporated. During flight, apart from efforts to maintain flight, insects need to maintain their metabolism at a rate higher than that required at rest (basal metabolic rate). The  respiration ratio ($R$) accounts for the increase in metabolism due to continuous flight. Only a fraction of the chemical power expended is converted into mechanical power, which is accounted for by the conversion efficiency ($\eta$). Combining all these factors final expression for chemical power is given by:
\begin{equation}
    P_{chem} = \frac{R(P_{mech}+P_{met})}{\eta}    
    \label{P_chem}
\end{equation}
where the metabolic power $P_{met}$ is defined as $P_{met} = {\eta}P_{bmr}$. Here $P_{bmr} \equiv 3.79 (m  - m_{fat})^{0.723}$. Here  $m_{fat}$ is the fat mass and $m_{fat} = 0.35m$ at $t=0$. 

\par
\textbf{Lift to Drag ratio}\\
Lift to drag ratio is a measure of distance covered by the insect per unit fuel energy consumed and can be related to the chemical power (Eq. \ref{L_D}).
\begin{equation}
N = \frac{mgV_t}{{\eta}P_{chem}}
\label{L_D}
\end{equation}

\par
\textbf{Wingbeat frequency}\\
The wingbeat frequency (wingbeats per second) is a measure of power available from flight muscles, that is generated by the contraction and expansion of the muscle during flapping. Based on dimensional analysis the  wingbeat frequency is correlated to the body mass (m), wing span (B), wing area (S), air density ($\rho$) and gravity (g).
To some degree, the wingbeat frequency is under the control of the insect, but usually it does not vary much from the natural wingbeat frequency, which  is determined by physics of beating wings (Eq. \ref{wbf})
\begin{equation}
    \textit{f} = m^{3/8}g^{1/2}B^{-23/24}S^{-1/3}{\rho}^{-3/8}
    \label{wbf}
\end{equation}

\textbf{Time marching computation}

 A MATLAB code was developed for the numerical solution of the energetics model. The time marching computation is performed to compute the morphological parameters and flight parameters at time intervals of 6 minutes. In order to obtain the instantaneous mass of the insect $m$, we require the rate of mass burnt, $dm/dt$. It is obtained using the mass burning relation, $dm/dt = P_{chem} / e$ where $e$ is the energy density of the fuel. We assume that the insect obtains the $95\%$ of the chemical power by burning fat and the rest from muscle mass consisting of protein. Therefore, $m_{fat}$ is updated after each time step till $m_{fat} \geq 0$. 
A constant specific work criterion (eq. \ref{sp_work}) was used to compute muscle-burning rate. 
\begin{equation}
q_m = \frac{P_{mech}}{m_{musc}(1-\zeta)\textit{f}}
\label{sp_work}
\end{equation}\\
where, $m_{musc}$ is flight muscle mass and $\zeta$ is Volume fraction of mitochondria in flight muscles. 

Substituting the expressions for $P_{mech}$ (eq. \ref{P_{mech}}), wingbeat frequency (eq. \ref{wbf}) and $m_{musc} \equiv 0.15m$ a fourth degree polynomial for specific work ($q_m$) is obtained in terms of instantaneous migration velocity. The  specific work is computed at $t= 0$ and set as a constant for the rest of the time marching procedure to solve for the migration velocity until $V \leq V_{mr}$; thereafter the migration occurs at $V_{mr}$.

The flowchart in Fig. \ref{migration_range} shows the algorithm used for migration range calculations. The model is validated with the results of \citet{pennycuick2008modelling} for the bird Great Knot (see SI sec. 1).
\begin{figure}[!hb]
    \begin{center}
\tikzstyle{startstop} = [rectangle, rounded corners, minimum width=3cm, minimum height=1cm,text centered,text width=3cm, draw=black, fill=red!30]
\tikzstyle{io} = [trapezium, trapezium left angle=70, trapezium right angle=110, minimum width=1cm, minimum height=1cm, text centered, text width=4cm, draw=black, fill=blue!30]
\tikzstyle{process} = [rectangle, minimum width=3cm, minimum height=1cm, text centered, text width=5cm, draw=black, fill=orange!30]
\tikzstyle{decision} = [diamond, minimum width=3cm, minimum height=2cm, text centered, text width=3cm, draw=black, fill=green!30]
\tikzstyle{datastore}= [rounded rectangle, rounded rectangle east arc=concave, rounded rectangle arc length=150, minimum width=3cm, minimum height=2cm, text centered, text width=3cm,draw=black, fill=cyan!30]
\tikzstyle{arrow} = [ultra thick,->,>=stealth]
\begin{tikzpicture}[node distance=2cm]
\node (start) [startstop]{Start};
\node (pc) [process, below of=start]{Power curve calculation};
\node (vmp) [io, below of=pc]{Setting initial speed as $V_{mp}$};
\node (ts) [process, below of=vmp]{Run for one time step};
\node (op) [io, below of=ts]{Power, specific work, and fuel consumption};
\node (pc1) [process, below of=op]{Power curve calculation for current composition};
\node (vel) [process, below of=pc1]{Velocity for next time step using specific work criteria if V $<$ $V_{mr}$ else V=$V_{mr}$ };
\node (run) [process, below of=vel]{Running the time step till $m_{fat}$=0};
\node(end) [startstop, below of=run]{End};

\draw [arrow] (start) -- (pc);
\draw [arrow] (pc) -- (vmp);
\draw [arrow] (vmp) -- (ts);
\draw [arrow] (ts) -- (op);
\draw [arrow] (op) -- (pc1);
\draw [arrow] (pc1) -- (vel);
\draw [arrow] (vel) -- (run);
\draw [arrow] (run) -- (end);
\end{tikzpicture}
\end{center}
    \caption{Flowchart of time marching computation for migration range calculation}
    \label{migration_range}
\end{figure}
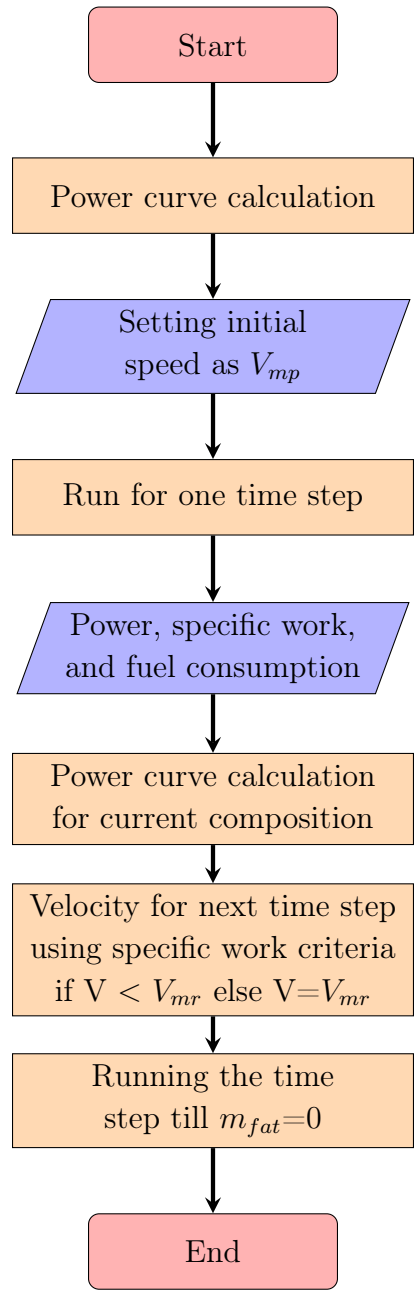


\subsubsection{Energetics of Pantala Flavescens}
 We captured dragonflies at the IIT Kharagpur campus to determine the input parameters for the DEM. We measured morphological parameters (mass, wing span, wing area, and frontal area), using weighing balance and vernier, and other relevant data like, aspect ratio ($AR = B^2/S$), fat mass ($m_{fat} = 0.85m$), and muscle mass ($m_{musc} = 0.15m$) were extracted using these parameters. Profile power constant and induced power factor for the calculations are identical to that of birds, as they were found to be satisfactorily applicable by \citet{May1991} and \citet{Azuma1988} in their calculations of the power curve for dragonflies and similar U-shaped power curve has been estimated for large insects like moths as well \citep{warfvinge2017power}. The body drag co-efficient $C_{db}$ was obtained from \citet{May1991} for dragonflies with similar mass. The DEM parameters and results are detailed in SI sec. 2 .
 
\subsection{Dragonfly Path planning model(DPM)}

\begin{figure}[!ht]
\begin{center}
\tikzstyle{startstop} = [rectangle, rounded corners, minimum width=3cm, minimum height=1cm,text centered,text width=3cm, draw=black, fill=red!30]
\tikzstyle{io} = [trapezium, trapezium left angle=70, trapezium right angle=110, minimum width=1cm, minimum height=1cm, text centered, text width=2.5cm, draw=black, fill=blue!30]
\tikzstyle{process} = [rectangle, minimum width=3cm, minimum height=1cm, text centered, text width=3cm, draw=black, fill=orange!30]
\tikzstyle{decision} = [diamond, minimum width=3cm, minimum height=2cm, text centered, text width=3cm, draw=black, fill=green!30]
\tikzstyle{datastore}= [rounded rectangle, rounded rectangle east arc=concave, rounded rectangle arc length=150, minimum width=3cm, minimum height=2cm, text centered, text width=3cm,draw=black, fill=cyan!30]
\tikzstyle{arrow} = [ultra thick,->,>=stealth]

\begin{tikzpicture}[node distance=4cm]
\node (start) [startstop]{Start};
\node (wind) [io, below of=start,xshift=0cm,yshift=1cm] {Wind velocity field};
\node (lat) [io, left of=wind,xshift=-2cm] {Latitude and Longitude of starting and end point};
\node (dem) [io, right of=wind,xshift=2cm] {DEM};
\node (grid) [process, below of=lat,yshift=2cm] {Grid};
\node (vw) [process, below of=wind,yshift=2cm] {Wind velocity($V_w$, {$\theta$})};
\node (vs) [process, below of=dem,yshift=2cm] {Dragonfly Velocity($V_s$)};
\node (dist) [io,below of=grid,yshift=1.5cm] {Distance, Grid direction({$\alpha$})};
\node (in1) [process, below of=vw,yshift=1.5cm] {Compute $\delta$ and $V_t$};
\node (time) [io,below of=in1,yshift=1.5cm] {Time};
\node (cm) [datastore,right of=time,xshift=1.5cm] {Generate and store time in cost matrix};
\node (DA) [process,below of=cm,xshift=0cm,yshift=1cm] {Dijkstra's Algorithm};
\node (R) [io,left of=DA,xshift=-1cm,yshift=0cm] {Optimal Route and time of flight};
\node (end) [startstop, left of=R,xshift=-1cm]{End};

\draw [arrow] (start) -| (lat);
\draw [arrow] (start) -- (wind);
\draw [arrow] (start) -| (dem);
\draw [arrow] (lat) -- (grid);
\draw [arrow] (grid) -- (dist);
\draw [arrow] (dist) -- (in1);
\draw [arrow] (wind) -- (vw);
\draw [arrow] (vw) -- (in1);
\draw [arrow] (dem) -- (vs);
\draw [arrow] (vs) |- (in1);
\draw [arrow] (dist) |- (time);
\draw [arrow] (in1) -- (time);
\draw [arrow] (time) -- (cm);
\draw [arrow] (cm) -- (DA);
\draw [arrow] (DA) -- (R);
\draw [arrow] (R) -- (end);
\end{tikzpicture}
\end{center}
    \caption{Flowchart for dragonfly path planning algorithm}
    \label{DPM}
\end{figure}
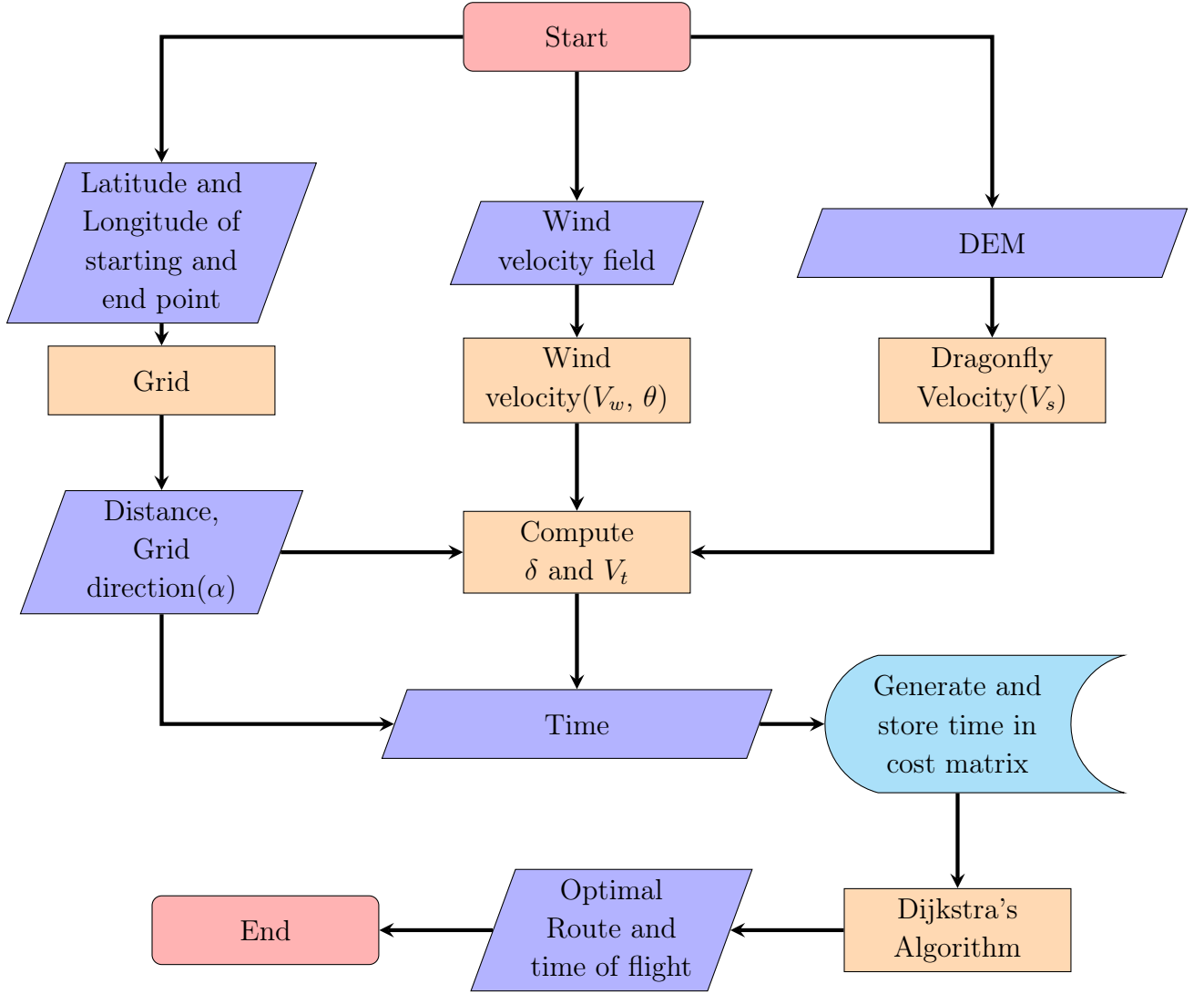

Dijkstra's algorithm \cite{dijkstra1959note} was used to evaluate the role of wind in the migration of P. flavescens. The algorithm searches for the shortest path between the starting  and the end point, given that at least one such path exists. We marked the the stopovers as the starting and the end points. An open-source MATLAB code of Dijkstra's algorithm \citep{DimasAryo2021} was developed further by modifying the cost matrix for computing the optimal path and the associated time. 
The combination of the cost matrix along with Dijkstra's algorithm is designated as the dragonfly path planning model(DPM) hereafter.\par
Either of the two optimization criteria, the time of flight and the distance covered, can be used for generating the cost matrix. However, the primary constraint during migration is the fuel reserve which places an upper limit on the time of flight, but the distance covered depends on the time of flight, the migration velocity of the insect and the local wind velocity. Therefore, the time of flight is a more fundamental constraint associated with the fuel reserve and hence chosen as the cost function \citep{warfvinge2017power}.\par
Three key inputs are required for the generation of the cost matrix: the dragonfly migration velocity, which in this case is $V_{mr}$ and is obtained from DEM; the local wind velocity obtained from NOAA \citep{NOAA} ; the global position in terms of latitude and longitude. We select the latitude and the longitude of the starting and end points while initializing the DPM. Thereafter the rectangular area between the starting and end point, with the diagonal as the geodesic distance between the two points, is discretized with 30  grid-points in each direction. A separate discretization grid is used for each transoceanic leg of the migration. 
Each node of the grid is assigned wind velocity, latitude and longitude, dragonfly migration velocity, and possible flight directions (track). We consider eight possible directions for each internal node (see Fig. \ref{fig:grid and vel schematic}(a)); the boundary nodes have fewer directions. Based on the latitude and longitude distance between any two nodes is calculated. Also, based on dragonfly migration speed ($V_s$), local wind velocity, and the track, a resultant velocity between two neighboring nodes is computed. The resultant velocity and the distance determine the time taken to travel between the two nodes that serves as the cost function between any two nodes. We compute the cost matrix using all possible combinations of nodes; that is the time taken to travel for each possible route constitutes the associated entry of the cost matrix. Using the cost matrix in Dijkstra's algorithm \parencite{dijkstra1959note}, we calculate the optimal time and the optimal route for each leg of the transoceanic migration. We also calculate the total distance and the fuel consumed from the optimal route. Fig.\ref{DPM} shows the flowchart for the DPM. 
\begin{figure}[h!]%
\begin{center}
\subfloat[]{\includegraphics[trim=50 120 50 80,clip,width=0.5\textwidth]{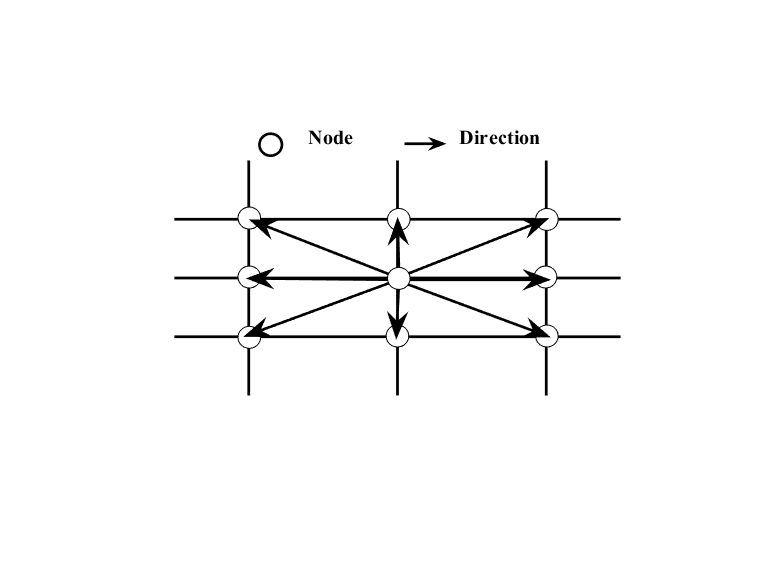}}\quad
\subfloat[]{\includegraphics[width=0.4\textwidth]{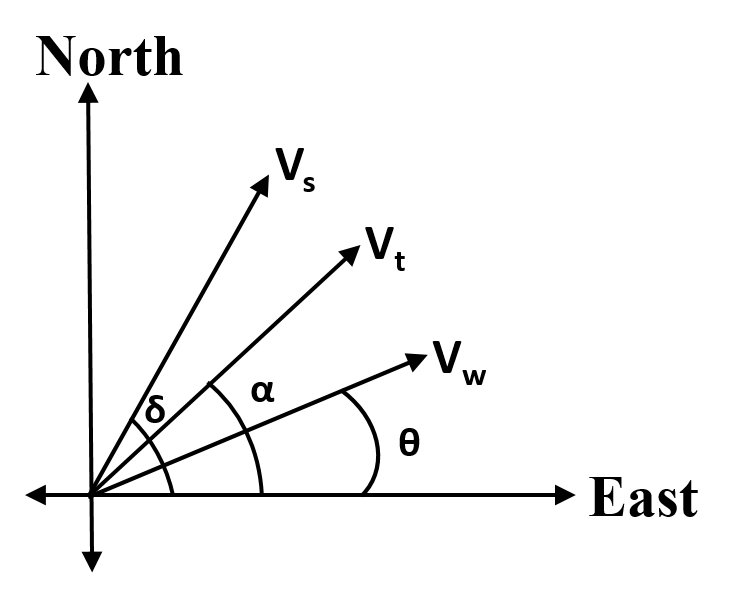}}
\caption{%
(a) A section of the grid showing all possible directions of movement from a node. (b) Schematic representation of dragonfly speed ($V_s$), wind speed ($V_w$), resultant speed ($V_t$), dragonfly heading ($\theta$), wind direction ($\delta$), and dragonfly track ($\alpha$). At any node $V_s$, $V_w$, $\theta$, and $\alpha$ are known, and $V_t$ and $\delta$ are calculated using the schematic.}%
\label{fig:grid and vel schematic}
\end{center}
\end{figure}

Here $V_w$ is the magnitude of wind velocity calculated from two planar components, u-wind (u component of wind velocity, positive in due east) and v-wind(v component of wind velocity, positive in due north). The magnitude of dragonfly velocity is $V_s$ (we assume $V_s = V_{mr}$), and $V_t$ is the magnitude of resultant velocity, $\theta$ is wind direction, $\alpha$ is dragonfly track, that is the direction relative to the ground and is fixed by the grid (see Fig. \ref{fig:grid and vel schematic}(a)). Here $\delta$ is dragonfly heading; the direction relative to the wind field that is required to maintain the track. All angles are measured with respect to due east (see Fig. \ref{fig:grid and vel schematic}(b)).
\subsection{Passive tracer trajectory}
We simulate the trajectory of a migrating dragonfly species, \textit{Pantala flavescens}, under the influence of the atmospheric wind field as if it acts as a passive tracer particle and gets purely convected by the wind field. We used wind data from NOAA \citep{NOAA} for the region of interest. We use wind at a height of 850 hPa, making the trajectory two-dimensional. We have developed a MATLAB script for solving the equations of motion (Eq. \ref{passive_tracer}) using the modified Euler method (further details of the numerical method are provided in \citet{pozrikidis2016fluid, giri2022colliding}). The particle at any instant assumes the local wind velocity ($u_x,u_y$) and the velocity is updated at each time step({$\Delta$}t = 60s) after advancing the particle location $X(t), Y(t)$ in time.
\begin{eqnarray}
\frac{dX}{dt}  = u_{x}(X(t), Y(t), t) & ; &
\frac{dY}{dt}  = u_{y}(X(t), Y(t), t)
\label{passive_tracer}
\end{eqnarray}
\end{justify}

\newpage
\section{Supplementary Information}
\subsection{Validation of computational model}
The energetics model developed is validated with the existing Flight model of \citet{pennycuick2008modelling}, for bird Great knot. The parameters used for the calculations are shown in Table \ref{table1}.

\begin{table}[h!]
\centering
\begin{tabular}{|c|c|c|c|}
\hline
\textbf{Parameter}&\textbf{Value}&\textbf{Parameter}&\textbf{Value}\\
\hline
Mass&0.233 kg&Wing span&0.586 m\\
\hline
Wing area&0.0397 $m^2$ &Aspect ratio&8.65\\
\hline
Frontal area&0.00308 $m^2$ &Drag co-efficient&0.1\\
\hline
Profile power constant&8.4 &Induced power factor&0.9\\
\hline
g&9.81 $m/s^2$ &$\rho$&0.909 $kg/m^3$\\
\hline
Fat fraction&0.385&Flight muscle fraction&0.144\\
\hline
Airframe fraction&0.472&Respiration ratio&1.1\\
\hline
Conversion efficiency&0.23&&\\
\hline
\end{tabular}
\caption{Basic parameters of Great Knot}
\label{table1}
\end{table}

\begin{figure}[h!]%
\begin{center}
\subfloat[]{\includegraphics[trim=400 10 400 20,clip,width=0.45\textwidth]{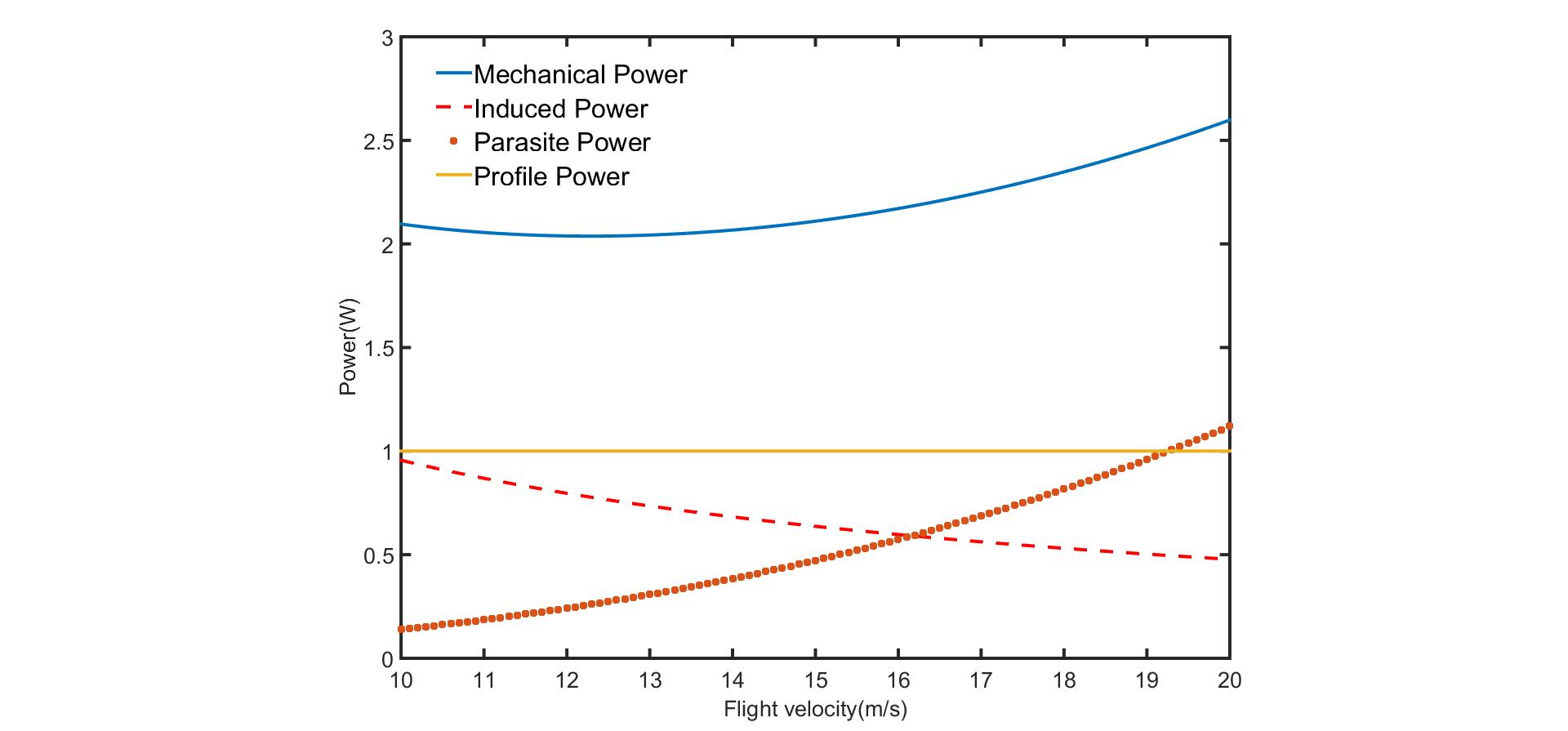}}\quad
\subfloat[]{\includegraphics[trim=400 10 400 20,clip,width=0.45\textwidth]{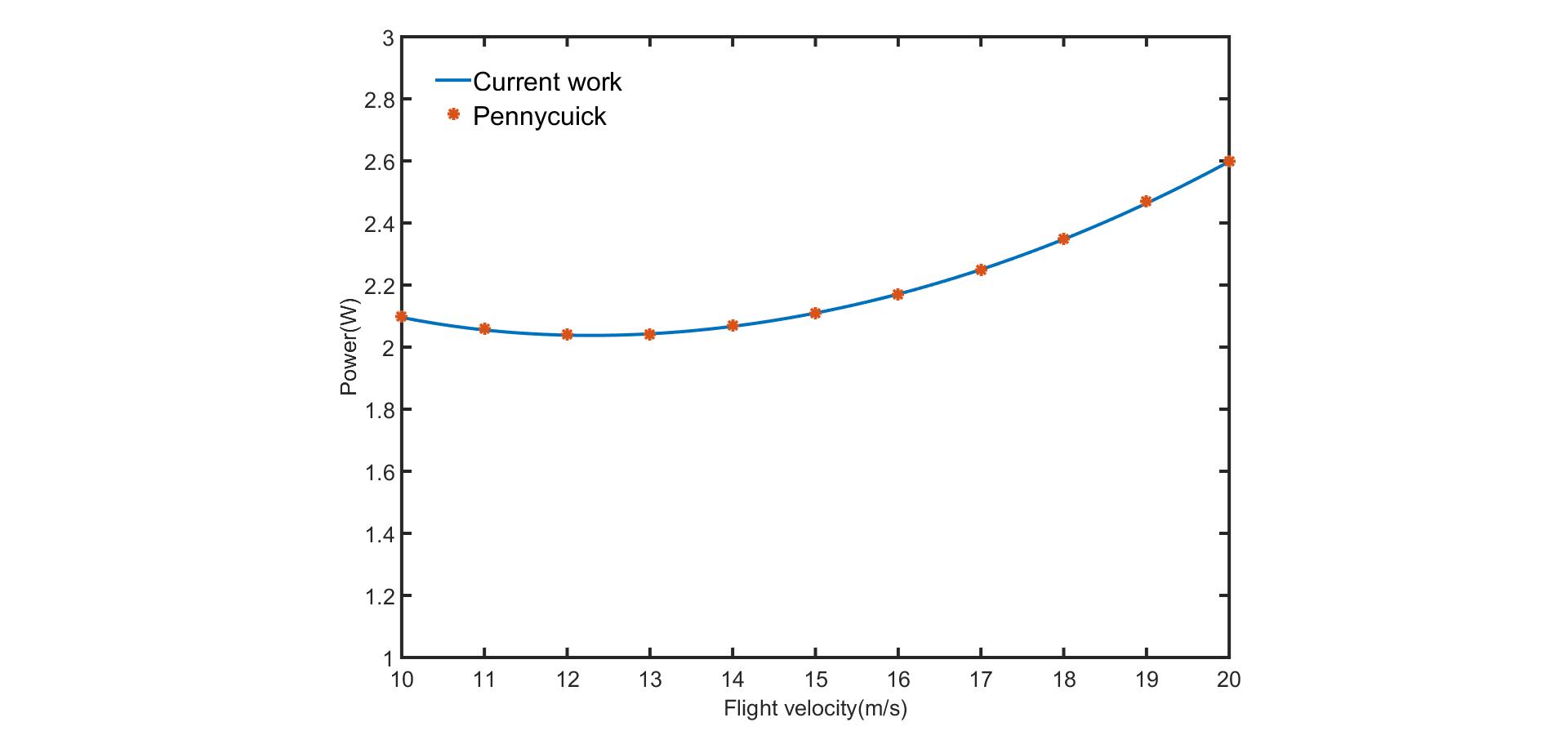}}
\caption{%
Power curve for Great Knot and power curve comparison for Great Knot}%
\label{fig:Power_curve_GK}
\end{center}
\end{figure}

\textbf{Power curve validation}\\
Power curve for Great knot is shown in Fig. \ref{fig:Power_curve_GK}(a) and comparison of two models is shown in Fig. \ref{fig:Power_curve_GK}(b) and Table \ref{table2}. From the comparison it is evident that the current numerical model is in agreement with the power curve of Great knot \parencite{pennycuick2008modelling}. 

\begin{table}[h]
      \centering
       \begin{tabular}{|c|c|c|}
\hline
\textbf{Parameter}&\textbf{Current work}&\textbf{\citet{pennycuick2008modelling}}\\
\hline
$V_{mp}$ (m/s)&12.30&12.30\\
\hline
$V_{mr}$ (m/s)&19.80&19.80\\
\hline
\end{tabular}
\caption{Power curve comparison for Great Knot}
\label{table2}
\end{table}
\newpage
\textbf{Migration validation}\\
Migration calculation results as compared to \citet{pennycuick2008modelling} are shown in Table \ref{Migration validation}. The results are in close agreement and validates the current model.
 
\begin{table}[h]
      \centering
       \begin{tabular}{|c|c|c|}
\hline
Parameter&Current work&\citet{pennycuick2008modelling}\\
\hline
Final mass&0.1111 kg&0.1111 kg\\
\hline
Range&8258 km&8464 km\\
\hline
Time of flight&132 h&135.4 h\\
\hline
\end{tabular}
\caption{Migration calculation comparison for Great Knot}
\label{Migration validation}
\end{table}

\begin{justify}
\subsection{Dragonfly migration energetics results}
The dragonfly energetics model(DEM) is used for the power curve calculation and migration calculation of \textit{P. flavescens}. Table \ref{PF parameters} shows the input data used for power curve and migration calculation. 
The power curve calculation and the migration calculation results are shown in Table \ref{Migration calculation of PF}.
\begin{table}[h!]
\centering
\begin{tabular}{|c|c|c|c|}
\hline
\textbf{Parameter}&\textbf{Value}&\textbf{Parameter}&\textbf{Value}\\
\hline
Mass&0.345 g&Wing span&0.079 m\\
\hline
Wing area&0.00146 $m^2$ &Aspect ratio&4.28\\
\hline
Frontal area&4.715$\times$ $10^{-5}$ $m^2$ &Drag co-efficient&0.4\\
\hline
Profile power constant&8.4&Induced power factor&0.9\\
\hline
g&9.81 $m/s^2$ &$\rho$&1.056 $kg/m^3$\\
\hline

Fat fraction&0.35&Flight muscle fraction&0.15\\
\hline
Airframe fraction&0.5&Respiration ratio&1.1\\
\hline
Conversion efficiency&0.23&&\\
\hline
\end{tabular}
\caption{Basic parameters of Pantala Flavescens}
\label{PF parameters}
\end{table}
\newpage
\begin{figure}[h!]%
\begin{center}
\includegraphics[trim=400 10 400 20,clip,width=0.5\textwidth]{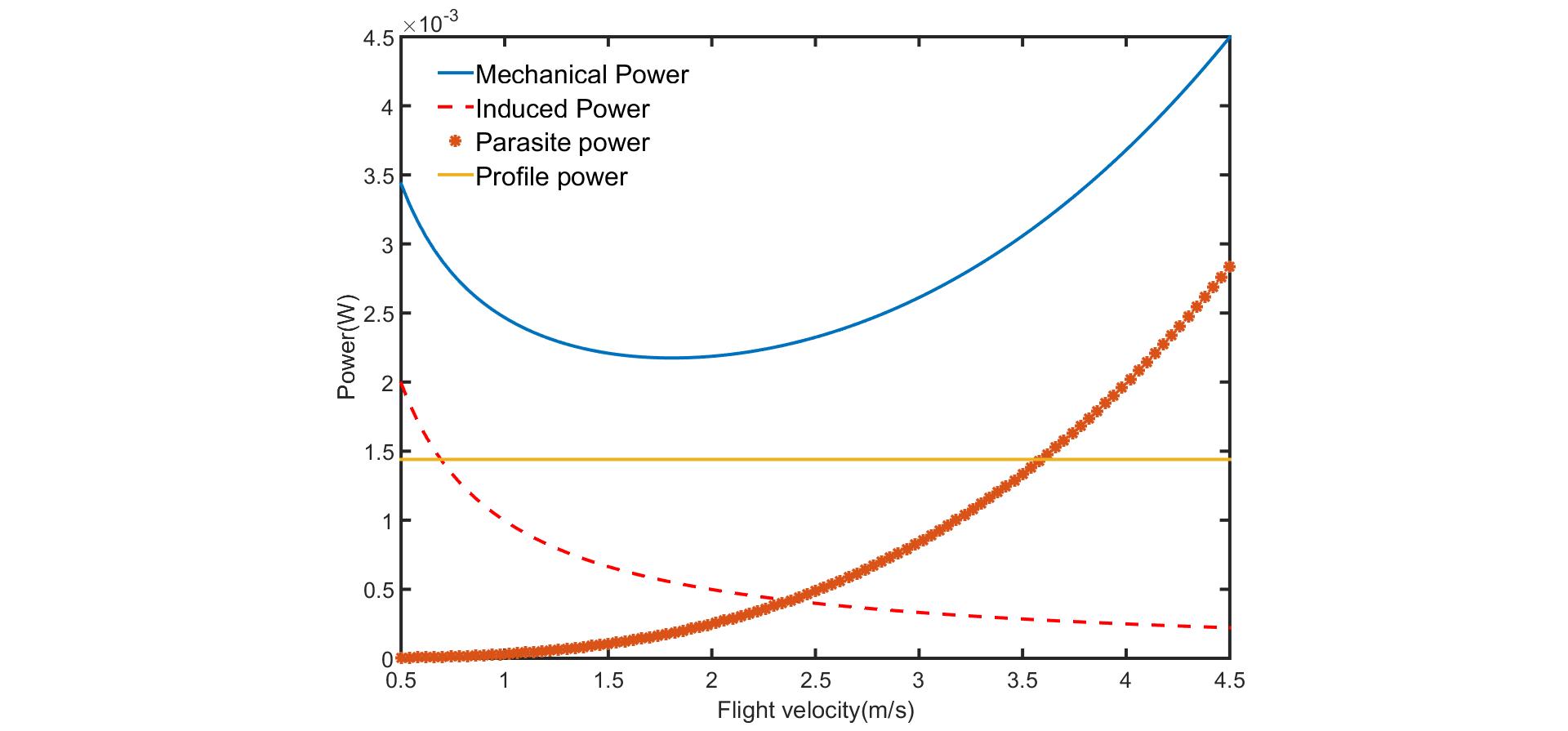}
\caption{%
Power curve for Pantala Flavescens}%
\label{fig:Power_curve_PF}
\end{center}
\end{figure}
\begin{table}[ht!]
\centering
\begin{tabular}{|c|c|c|c|c|c|}
\hline
Parameter&$V_{mp}$&$V_{mr}$&Consumed mass&Range&Time of flight\\
\hline
Value&2.42 m/s&4.5 m/s&0.1643 g&1400 km&90 h\\
\hline
\end{tabular}
\caption{Migration calculation results for Pantala Flavescens}
\label{Migration calculation of PF}
\end{table}
\begin{table}[h!]
\centering
\begin{tabular}{|c|c|c|}
\hline
\textbf{Place}&\textbf{Reference}&\textbf{Months (Remarks)}                                           \\ \hline

\multirow{2}{*}{Maldives}   & \citet{Anderson2009} & Oct-Dec \\ \cline{2-3}
                          & \citet{olsvik1992dragonfly} & Nov(Mating) \\ \hline
\multirow{5}{*}{Seychelles}   & \citet{bowler2003odonata} & Nov \\ \cline{2-3}
                          & SBRC & Dec-Jan \\ \cline{2-3}
                          & alphonse-island.com & Mar \\ \cline{2-3} 
                          & \citet{wain1999odonata} & Nov(Breeding)\\ \cline{2-3}
                          & \citet{campion1913no} & Nov(Location: Aldabra)\\ \cline{2-3}
                          & \citet{samways1998establishment,samways2010tropical} & Nov-Apr \\ \hline 
South Africa&\citet{samways1998divergence}& Dec-Feb (breeding) \\ \hline
\multirow{2}{*}{Mozambique/Malawi}   & \citet{dijkstra2004dragonflies} & Nov and Jan \\ \cline{2-3}
                       & \citet{bernard2020new} & Dec/Mar-Apr \\ \hline 
Tanganyika&\citet{bartenef1931uber}&Dec-Jan\\ \hline 
Uganda&\citet{bartenef1931uber}&Mar-Apr,Oct\\ \hline 
India&\citet{fraser1924survey}&Sep-Nov (Departure on \\ &&  'annual migration')\\ \hline 
Amsterdam Island&\citet{devaud2019first}& Feb\\ \hline
Chagos Archipalego&\citet{carr2022odonata}& Oct-Nov \\ \hline
\end{tabular}
\caption{Sighting at different locations reported in the literature}
\label{tab:sighting}
\end{table}
\begin{table}[h!]
\centering
\begin{tabular}{|c|c|c|c|}
\hline
Section&Time(hr)&Section&Time(hr)\\
\hline
S1 - S6&115&S2 - S6&66\\
\hline
S1 - S2&52&S2 - S5&37.5\\
\hline
S1 - S3&88&S3 - S6&43\\
\hline
S1 - S4&85&S4 - S6&31\\
\hline
S1 - S5&71&S5 - S6&49\\
\hline
\end{tabular}
\caption{Time to cover different section branching network of various possibilities of reaching Maldives (S6) from  Cherrapunji (S1) with stopovers at Visakhapatnam [17.5N 82.5E](S2), Mangalore[13N 75E] (S3), Thiruvananthapuram[9N 77E] (S4) and in Sri Lanka[9N 81E] (S5).}
\label{tab:network time}
\end{table}
\end{justify}

\newpage
\begin{justify}
\printbibliography[
heading=bibintoc,
title={References}]
\end{justify}

\end{document}